\documentclass[12pt]{iopart}
\usepackage{iopams} 
\expandafter\let\csname equation*\endcsname\relax
\expandafter\let\csname endequation*\endcsname\relax
\usepackage{amsmath}
\usepackage{amsfonts,amssymb}
\usepackage{amscd}
\usepackage{graphics,epsfig,subfigure}
\usepackage{color,bm}
\usepackage{hyperref}
\begin{document}
\title{Independent-oscillator model and the quantum Langevin equation for an oscillator: A review}
\author{Aritra Ghosh\footnote{E-mail: ag34@iitbbs.ac.in} and Malay Bandyopadhyay\footnote{E-mail: malay@iitbbs.ac.in}}
\address{School of Basic Sciences, Indian Institute of Technology Bhubaneswar, Jatni, Khurda, 752050, India}
\author{Sushanta Dattagupta\footnote{E-mail: sushantad@gmail.com}}
\address{National Institute of Technology, Mahatma Gandhi Road, Durgapur, 713209, India\\
and,\\
Sister Nivedita University,
New Town, Block DG 1/2, Action Area I, 
Kolkata, 700156, India}
\author{Shamik Gupta\footnote{E-mail: shamikg1@gmail.com}}
\address{Department of Theoretical Physics, Tata Institute of Fundamental Research, 1, Homi Bhabha Road,
Mumbai, 400005, India}

\vspace{0.25cm}
\begin{indented}
\item[]May 2024
\end{indented}
\tableofcontents

\begin{abstract}
This review provides a brief and quick introduction to the quantum Langevin equation for an oscillator, while focusing on the steady-state thermodynamic aspects. A derivation of the quantum Langevin equation is carefully outlined based on the microscopic model of the heat bath as a collection of a large number of independent quantum oscillators, the so-called independent-oscillator model. This is followed by a discussion on the relevant `weak-coupling' limit. In the steady state, we analyze the quantum counterpart of energy equipartition theorem which has generated a considerable amount of interest in recent literature. The free energy, entropy, specific heat, and third law of thermodynamics are discussed for one-dimensional quantum Brownian motion in a harmonic well. Following this, we explore some aspects of dissipative diamagnetism in the context of quantum Brownian oscillators, emphasizing upon the role of confining potentials and also upon the environment-induced classical-quantum crossover. We discuss situations where the system-bath coupling is via the momentum variables by focusing on a gauge-invariant model of momentum-momentum coupling in the presence of a vector potential; for this problem, we derive the quantum Langevin equation and discuss quantum thermodynamic functions. Finally, the topic of fluctuation theorems is discussed (albeit, briefly) in the context of classical and quantum cyclotron motion of a particle coupled to a heat bath.
\end{abstract}

\section{Introduction}
Coupling a system to its environment turns it into a dissipative or open system. Open systems are ubiquitous in nature, from a cup of freshly-brewed coffee that continuously loses heat to the environment, to complex biological processes inside living cells. A paradigm for open systems is provided by the problem of Brownian motion, named after Robert Brown. The observation of the Brownian motion has been by Brown himself in 1827 in the context of pollen grains (the `Brownian' particles) suspended in a liquid in thermal equilibrium at temperature $T$ and being constantly buffeted by the liquid molecules, resulting in a dissipative dynamics of the Brownian particles~\cite{brown}. It should be remarked that prior to Brown, Ingenhousz in 1784 reported the erratic movement of coal dust on the surface of alcohol \cite{janI} (see also, Ref.~\cite{Hanggirev}). A satisfactory microscopic understanding of the Brownian motion however emerged much later, in the 20th century with the seminal work of Einstein~\cite{brown1}. Subsequently, many
researchers made notable contributions to this field, and some significant outcomes include the Einstein-Smoluchowski relation~\cite{brown2}, the Langevin equation~\cite{brown3}, and the Fokker-Planck-equation description of the Brownian motion~\cite{brown4}, as well as extensions and applications of the Brownian motion, i.e., the Ornstein-Uhlenbeck (OU) process~\cite{brown5}, the Kramers problem~\cite{kramer}, the phenomenological stochastic model for lineshapes in spectroscopy~\cite{lineshape}, and many others. The work of Einstein is path breaking in that it provided the
 first formal connection between dissipative forces and 
thermal fluctuations, encoded in the so-called Einstein relation that relates the diffusion constant $D$ of the Brownian particles, a measure of fluctuations in their motion as a function of time, to the strength of dissipation $\gamma$ in the dynamics of the Brownian particles and the amount of thermal fluctuations ($=k_BT$, with $k_B$ the Boltzmann constant). This connection between dissipation and related fluctuations was later put on a solid foundation by Johnson and Nyquist in the context of thermal agitation of charge carriers inside electrical conductors in equilibrium~\cite{johnson1,johnson2,nyquist}. Relations between the dissipative and fluctuating aspects of the dynamics of an open system are clubbed together under fluctuation-dissipation relations. 

Quantum mechanically, the problem is more subtle. It is natural to seek a description in terms of a quantum system (say, a single particle) in contact with a heat bath such that the overall dynamics of the system-plus-heat-bath is unitary. Then, the total Hamiltonian describing the situation will be of the form \(H = H_\mathrm{S} + H_\mathrm{B} + H_\mathrm{SB}\), where \(H_\mathrm{S}\) and \(H_\mathrm{B}\) are respectively the Hamiltonian operators of the system and the bath, while \(H_\mathrm{SB}\) summarizes the interaction between them. In the Schr\"{o}dinger picture, the time evolution of the state describing the system and the bath together is of the form
\begin{equation}\label{TDSE}
\mathrm{i}\hbar \frac{\partial}{\partial t} |\Psi (t) \rangle = \big( H_\mathrm{S} + H_\mathrm{B} + H_\mathrm{SB} \big)  |\Psi (t) \rangle,
\end{equation}
where $|\Psi(t)\rangle$ denotes the state of the composite system, and $\hbar$ denotes the reduced Planck's constant.
However, one is typically interested in the dynamics of the system alone, and it is not immediately clear from Eq.~(\ref{TDSE}) how to appropriately eliminate or `trace out' the bath variables so that what remains is the information about the dynamics of the `open' quantum system \(\mathrm{S}\). In a seminal work~\cite{FV}, Feynman and Vernon suggested the use of path integrals to trace out the environmental effects, i.e., the degrees of freedom constituting the bath. The procedure involves the computation of a quantity known as the `influence functional', which appropriately describes the effect of the bath on the system. Caldeira and Leggett~\cite{CL,CL1,CL2} extended the Feynman-Vernon scheme to the case of a quantum particle moving in an arbitrary potential, and in contact with a bath of quantum oscillators. Utilizing the path integral formalism, Caldeira and Leggett worked with the density operator in the Schr\"{o}dinger-picture representation. Following the work of Feynman and Vernon~\cite{FV}, it was demonstrated in a celebrated paper by Ford {\it et al.}~\cite{lho1} that a description of the heat bath as an infinite collection of harmonic oscillators is indeed a good one (both classically and quantum mechanically), in the sense that in the Heisenberg picture, the dynamical operators of the system \(\mathrm{S}\) admit a Brownian motion-like dynamics when the equations of motion for the bath variables are solved and are substituted for into the former. Stochastic (noise) terms depending upon the initial conditions of the bath variables arise in the equation(s) of motion describing the system, and one assumes that these initial conditions are distributed according to some
statistical distribution such as the Boltzmann distribution so as to model the situation that the bath is in equilibrium all through the time evolution of the system. In this situation, one indeed gets a quantum (operator-valued) Langevin equation describing the quantum Brownian motion of the system. 
Over the years, there have been several works within this framework, which are devoted to study of a quantum harmonic oscillator (constituting the system of interest $\mathrm{S}$) in contact with a heat bath consisting of an infinite set of quantum harmonic oscillators~\cite{lho2,lho3,lho4,lho5,lho6,lho7}. 

The main purpose of this paper is to briefly and quickly introduce some aspects of the active area of quantum Brownian motion, via the simple example of a system comprising a quantum harmonic oscillator that is coupled to a heat bath composed of an infinite number of independent quantum harmonic oscillators. We shall follow the route due to Ford {\it et al.}, i.e., the approach involving the quantum Langevin equation describing the reduced dynamics of the system that is in contact with a heat bath, particularly stressing upon the stationary state and the thermodynamic aspects~\cite{quantumreg,fordphysicae2005,ocon1,15,FordQT,hanggi,PRE79,specificheatingold,sdg2,18,20}. Although it may come as a surprise that thermodynamics can be consistently formulated for a single particle, which is far from the thermodynamic limit that one invokes in traditional thermodynamic studies, it turns out that not only does a thermal description follow naturally from the framework of classical and quantum Brownian motion, it is also consistent with all the laws of thermodynamics, including the third law~(see, for example, Refs. \cite{fordphysicae2005,ocon1,15,20}). We will contrast the expressions of average energy between classical and quantum Brownian oscillators, which will lead us to the recently-proposed quantum counterpart of energy equipartition theorem~\cite{jarzy1,jarzy2,jarzy3,jarzy4,jarzy5,jarzy6,kaur,kaur2,kaur1,ghoshelec,kaur3}. As an application of the framework relevant to condensed-matter physics, we shall discuss dissipative diamagnetism~\cite{kaur3,sdg1} and emphasize on the role of confining potentials. Following this, we will discuss a situation where the system (the oscillator) and the bath are connected via momentum-dependent couplings \cite{Leg,Andiss,Andiss0,Andiss1}; the specific case of a gauge-invariant model of a three-dimensional oscillator coupled to a heat bath via momentum variables in the presence of a vector potential is carefully analyzed \cite{momentum,momentum1}. Finally, we discuss the timely topic of fluctuation theorems \cite{GaCoFT,JaFT,JKFT,iso1,iso2,iso3,GSA_SDG_PRE,GSA_SDG,meas1,meas2,meas3,meas4}, focusing on work fluctuations in the context of dissipative quantum oscillators, for such relations are intimately related to the second law of thermodynamics. As it turns out, for simple situations such as the Brownian motion as described by a linear Langevin equation which has a built-in noise term, the fluctuation theorem takes the simple form of the Gallavotti-Cohen kind~\cite{GaCoFT,JKFT,GSA_SDG_PRE}. We analyze this for the case of the dissipative quantum cyclotron motion of a free particle and also its classical counterpart \cite{GSA_SDG_PRE}.

The review is divided into several sections. In the next section [Section~(\ref{qbmsec})], we introduce the quantum-mechanical problem of Brownian motion of a harmonic oscillator, and discuss various intricate details of the formalism based on the quantum Langevin equation. Sections~(\ref{thermsec}) and~(\ref{fesec}) summarize thermodynamic aspects of the dissipative quantum oscillator, with appropriate comparisons done with the classical case. We discuss the mean energy, the free energy, and the entropy of a quantum oscillator executing Brownian motion. The problem of dissipative diamagnetism is analyzed in Section~(\ref{ddsec}) in some detail. In Section~(\ref{mmcsec}), we describe the notion of momentum-dependent system-bath couplings, and focus on a dissipative quantum oscillator with momentum-momentum coupling between the system and the heat bath. Finally, we discuss about fluctuation theorems by taking the example of dissipative cyclotron motion in Section~(\ref{FluctuationTheoremSec}). We end with some discussion in Section~(\ref{remarkssec}).

\section{Quantum Langevin equation}\label{qbmsec}
We begin with the full Hamiltonian of the system-plus-bath, which writes as~\cite{FV,CL,CL1,CL2,lho1,lho6,ford1988,Weiss,ohmicdiverge}:
\begin{equation}\label{Htot}
H = \frac{p^2}{2m} + \frac{m \omega_0^2 x^2}{2} + \sum_{j=1}^N \Bigg[ \frac{p_j^2}{2 m_j} + \frac{m_j \omega_j^2}{2} \bigg(q_j - \frac{c_jx}{m_j \omega_j^2} \bigg)^2\Bigg],
\end{equation}
wherein one can identify the first two terms on the right hand side as being the Hamiltonian of our system S involving a one-dimensional quantum harmonic oscillator; the remaining terms describe the heat bath B and the system-bath (SB) coupling, the latter being of the bilinear kind:
\begin{eqnarray}
&&H_\mathrm{S}=\frac{p^2}{2m}+ \frac{m \omega_0^2 x^2}{2},\label{HS}\\
&&H_\mathrm{B}=\sum_{j=1}^{N} \Bigg[\frac{p_j^2}{2m_j}+\frac{1}{2}m_j\omega_j^2q_j^2\Bigg],\label{HB}\\
&&H_\mathrm{SB}=-x\sum_{j=1}^Nc_j q_j+x^2\sum_{j=1}^N\frac{c_j^2}{2m_j\omega_j^2}. \label{HSB}
\end{eqnarray}
Here, \(x\) and \(p\) are the position and momentum operators of the system, \(q_j\) and \(p_j\) are the corresponding operators of the \(j\)-th heat-bath oscillator, and the real constant $c_j$ denotes the coefficient of linear coupling between the coordinate of the particle and that of the $j$-th oscillator. The relevant commutation relations for the various position and momentum operators are as follows:
\begin{equation}
[x,p]=\mathrm{i}\hbar,~~~~~~~[q_j,p_k]=\mathrm{i}\hbar \delta_{jk},
\label{canonical-commutation-0}
\end{equation}
where $\delta_{jk}$
is the Kronecker-$\delta$ function.

It should be remarked that the coupling of the system to the heat bath leads to a certain bath-induced (negative) frequency shift (see, for instance, Refs. \cite{CL1,CL2,Leg,HL} for a detailed discussion). The purpose of adding the term $\mathcal{V}_{\rm r}(x) = \sum_{j=1}^{N} c_j^2x^2/(2m_j\omega_j^2)$ in the Hamiltonian~(\ref{Htot}) is to precisely cancel such a shift (this additional term acts as a counter-term\footnote{In general, whether one wants to add counter-terms to cancel such shifts depends upon the situation one is interested in. For instance, when coupling an atom to an electromagnetic field, it would not make sense to correct for the Lamb shift in the model.}); the system feels an `effective' potential of the form [(\ref{app11})] \(\mathcal{V}_{\rm eff}(x) = (\omega_0^2 - (\delta \omega)^2) m x^2/2\), which resembles an inverted-oscillator potential if \((\delta \omega) > \omega_0\). The energies are then not bounded from below and there is no ground state~\cite{Leg,ford1988,Weiss}. Therefore, one may write \(\mathcal{V}_{\rm eff}(x) = \frac{m \omega_0^2 x^2}{2} + \mathcal{V}_{\rm induced}(x)\), where \(\frac{m \omega_0^2 x^2}{2}\) is the bare potential, i.e., the potential experienced by the particle in the absence of the heat bath, while \(\mathcal{V}_{\rm induced}(x)\) is the bath-induced shift; it turns out that the potential \(\mathcal{V}_{\rm r}(x)\) is chosen as \(\mathcal{V}_{\rm r}(x) = - \mathcal{V}_{\rm induced}(x)\) (see Refs. \cite{CL1,CL2} and (\ref{app11}) for more technical details). Thus, the linear-coupling model is repaired by adding the counter-term $\mathcal{V}_{\rm r}(x) = \sum_{j=1}^{N} c_j^2x^2/(2m_j\omega_j^2)$ to the Hamiltonian. It is noteworthy that the linear-coupling model is also known as the Ullersma model \cite{lho4}, while that with the inclusion of the counter-term is known as the independent-oscillator model \cite{lho1,lho6,ford1988}. The latter coincides with the celebrated Caldeira-Leggett model \cite{CL,CL1,CL2,Leg}. While our system is in contact with a single heat bath, a recent work~\cite{Tong:2023} has considered the case of multimode Brownian oscillators in nonequilibrium scenarios with multiple reservoirs at different temperatures as a model for open quantum systems in the nonequilibrium regime.

Considering the total Hamiltonian~(\ref{Htot}), one may obtain the Heisenberg equations of motion for the system and the bath variables, namely, \(\mathrm{i} \hbar \mathrm{d}A/\mathrm{d}t = [A,H]\), for \(A = x,p,q_j,p_j\). One gets 
\begin{eqnarray}
\dot{x}(t) &=& \frac{[x,H]}{\mathrm{i} \hbar} = \frac{p(t)}{m},  \\
 \dot{p}(t) &=& \frac{[p,H]}{\mathrm{i} \hbar} = - m \omega_0^2 x(t) + \sum_{j=1}^N c_j \bigg(q_j(t) - \frac{c_j x(t)}{m_j \omega_j^2}\bigg), \\
\dot{q}_j(t) &=& \frac{[q_j,H]}{\mathrm{i} \hbar} = \frac{p_j(t)}{m_j}, \label{qjoperatoreqn}\\
\dot{p}_j(t) &=& \frac{[p_j,H]}{\mathrm{i} \hbar} =  - m_j \omega_j^2 \bigg(q_j(t) - \frac{c_j x(t)}{m_j \omega_j^2}\bigg). \label{pjoperatoreqn}
\end{eqnarray}
This yields
\begin{equation}\label{dotxqlederivation}
m\ddot{x}(t) + m \omega_0^2 x(t) = \sum_{j=1}^N c_j \bigg(q_j(t) - \frac{c_j x(t)}{m_j \omega_j^2}\bigg),
\end{equation}
\begin{equation}\label{qjsecondoperator}
\ddot{q}_j(t) + \omega_j^2 q_j(t) = \frac{c_j}{m_j} x(t). 
\end{equation}
The second equation above may be solved to get
\begin{eqnarray}
&&q_j(t)- \frac{c_jx(t)}{m_j\omega_j^2}=\Big(q_j(0)-\frac{c_jx(0)}{m_j\omega_j^2}\Big)\cos(\omega_jt)\nonumber \\&&+\frac{p_j(0)}{m_j\omega_j}\sin(\omega_jt)+\frac{c_j}{m_j\omega_j^2}\int_{0}^{t}\mathrm{d}t^{\prime}~ \dot{x}(t^{\prime})\cos[\omega_j(t-t^{\prime})]. \label{qjsol}
\end{eqnarray}
Plugging Eq.~(\ref{qjsol}) into Eq.~(\ref{dotxqlederivation}), one gets
\begin{eqnarray}
&&m\ddot{x}(t) + m \omega_0^2 x(t)=-\int_{0}^{t}\mathrm{d}t^{\prime}~\dot{x}(t^{\prime})\sum_{j=1}^{N}\frac{c_j^2}{m_j\omega_j^2}\cos[\omega_j(t-t^{\prime})]\nonumber \\&&+\sum_{j=1}^{N}c_j \Big(q_j(0)-\frac{c_jx(0)}{m_j\omega_j^2}\Big)\cos(\omega_jt)+\sum_{j=1}^N\frac{c_jp_j(0)}{m_j\omega_j}\sin(\omega_jt) .
\end{eqnarray}
The above equation may be rewritten in the form of the quantum Langevin equation (see, for example, Refs.~\cite{ford1988,Das:2020}):
\begin{equation}\label{Eq.m}
m \ddot{x}(t) + \int_{0}^{t} \mathrm{d}t'~\mu(t - t') \dot{x}(t')+ m \omega_0^2 x(t) = f(t),
\end{equation}
where the noise term is given by
 \begin{equation}\label{noiseclassicalmicro}
 f(t)= \sum_{j=1}^{N}\Bigg[c_j \Big(q_j(0)-\frac{c_jx(0)}{m_j\omega_j^2}\Big)\cos(\omega_jt)+\frac{c_jp_j(0)}{m_j\omega_j}\sin(\omega_jt)\Bigg],
 \end{equation} which depends on the initial coordinates and momenta of the heat-bath oscillators (as well as on the initial position of the particle). Further, the friction kernel $\mu(t)$ is given by 
 \begin{equation}
 \label{classicalmut}
 \mu(t) =\Theta(t)\sum_{j=1}^{N}\frac{c_j^2}{m_j\omega_j^2}\cos(\omega_jt),
 \end{equation}
 where $\Theta(t\le 0)=0$ and $\Theta(t> 0)=1$ is the Heaviside step function. 
 
 \subsection{Initial-slip term}
 Now we elucidate the appearance of the so-called `initial-slip term', which arises when one attempts to derive the quantum Langevin equation from the independent-oscillator model, making the derivation of the quantum Langevin equation somewhat non-trivial. In fact, the noise operator \(f(t)\) as defined in (\ref{noiseclassicalmicro}) is not a stationary Gaussian process when averaged over the `bare' heat-bath distribution. To fully specify the reduced dynamics of the dissipative system, one needs to carefully specify the initial preparation of the system and the bath, as we discuss below.

 One would expect that initially, the system and the bath are decoupled from one another, i.e., we choose initial conditions of the system according to some rather suitable distribution function, while for the initial conditions of the heat-bath oscillators, we have a canonical distribution:
  \begin{equation}
\label{dmclass1111}
\rho_B(0) =\frac{\exp\left(- H_\mathrm{B}/k_BT\right)}{Z_{\rm B}},
\end{equation}
with 
 \begin{equation}
H_\mathrm{B}=\sum_{j=1}^N \left[\frac{p^2_j(0)}{2m_j}+\frac{1}{2}m_j\omega_j^2q^2_j(0)\right].
 \label{HBist}
 \end{equation}
 This amounts to taking the system and the bath to be decoupled at \(t = 0\), while the interactions are switched on for \(t \geq 0\). At thermal equilibrium, the system attains the same temperature as the bath. While the above-mentioned setting appears to be the most natural initial preparation, it does lead to the notion of an initial-slip term in the quantum Langevin equation that now follows from it \cite{bez,physicaA,hist}. Let us notice that because (\ref{dmclass1111}) dictates the initial distribution of the bath, one obtains \( \left\langle q_j(0)\right\rangle_{\rho_B(0)} =  \left\langle p_j(0)\right\rangle_{\rho_B(0)} = 0\) and the following second moments:
\begin{eqnarray}
    \left\langle q_j^2(0)\right\rangle_{\rho_B(0)} &=& \frac{{\rm Tr} \big[  q^2_j(0) e^{-H_{\rm B}/k_B T} ]}{{\rm Tr} \big[ e^{-H_{\rm B}/k_B T} ]}  =\frac{\hbar}{2m_j\omega_j}\coth \bigg(\frac{\hbar \omega_j}{2 k_B T}\bigg), \label{avgsqm211}
\end{eqnarray}
  \begin{eqnarray}
  \langle p^2_j(0)\rangle_{\rho_B(0)} &=& \frac{{\rm Tr} \big[ p_j^2(0) e^{-H_{\rm B}/k_B T} ]}{{\rm Tr} \big[ e^{-H_{\rm B}/k_B T} ]} = \frac{\hbar m_j \omega_j}{2} \coth \bigg(\frac{\hbar \omega_j}{2 k_B T}\bigg), \label{avgsqm311}
\end{eqnarray} and 
\begin{eqnarray}
    \left\langle q_j(0) p_j(0)\right\rangle_{\rho_B(0)} = - \left\langle  p_j(0) q_j(0)\right\rangle_{\rho_B(0)} = \frac{{\rm Tr} \big[  q_j(0) p_j(0) e^{-H_{\rm B}/k_B T} ]}{{\rm Tr} \big[ e^{-H_{\rm B}/k_B T} ]} = \frac{\mathrm{i}\hbar}{2}, \nonumber \\ \label{avgsqm411}
\end{eqnarray}
which implies that the noise as defined in (\ref{noiseclassicalmicro}) is not a Gaussian noise with zero mean with respect to \(\rho_\mathrm{B}(0)\). Instead, if we define
 \begin{equation}\label{gtnoise}
 g(t) \equiv  \sum_{j=1}^{N}c_j \Bigg[ q_j(0)\cos(\omega_jt)+\frac{p_j(0)}{m_j\omega_j}\sin(\omega_jt)\Bigg],
 \end{equation} then \(g(t)\) can be interpreted as a random noise which is Gaussian in nature. Looking at (\ref{noiseclassicalmicro}), (\ref{classicalmut}), and (\ref{gtnoise}), we may write the quantum Langevin equation (\ref{Eq.m}) as
 \begin{equation}\label{Eq.m112}
 m \ddot{x} + \int_{0}^{t} \mathrm{d}t'~\mu(t - t') \dot{x}(t')+ m \omega_0^2 x(t) + \mu(t) x(0) = g(t),
\end{equation} which holds on to an unphysical term \(\mu(t) x(0)\), known as the initial-slip term. The way out of this difficulty is to consider a setting where the system and the heat bath are in a state of thermal equilibrium even at the initial instant, i.e., at \(t = 0\), and therefore we consider a joint (conditional) distribution of the set $\{q_j(0),p_j(0)\}$ of the initial coordinates and the momenta of the heat-bath oscillators as given by the canonical-equilibrium distribution: 
 \begin{equation}
\label{dmclass}
\rho_\mathrm{B+SB}(0)=\frac{\exp\left(- H_\mathrm{B+SB}/k_BT\right)}{Z_{\rm B + SB}},
\end{equation}
with 
 \begin{equation}
H_\mathrm{B+SB}=\sum_{j=1}^N \left[\frac{p^2_j(0)}{2m_j}+\frac{1}{2}m_j\omega_j^2\left(q_j(0)-\frac{c_jx(0)}{m_j\omega_j^2}\right)^2\right].
 \label{HBSB}
 \end{equation}
Note that this scenario of introducing an ensemble of initial conditions is consistent with choosing the retarded solution~(\ref{qjsol}), in which time-reversal symmetry is explicitly broken, thereby introducing irreversibility into the problem. Corresponding to such an ensemble, one finds that the function $f(t)$ is a Gaussian noise, whose statistical properties may be deduced from the following averages:
 \begin{eqnarray}
   && \left\langle\left(q_j(0)-\frac{c_j x(0)}{m_j \omega_j^2}\right)\right\rangle_{\rho_\mathrm{B+SB}(0)} = \frac{{\rm Tr} \big[ \left(q_j(0)-\frac{c_j x(0)}{m_j \omega_j^2}\right) e^{-H_{\rm B +SB}/k_B T} ]}{{\rm Tr} \big[ e^{-H_{\rm B +SB}/k_B T} ]} = 0,     \label{avgsqm0} \\
   &&  \langle p_j (0)\rangle_{\rho_\mathrm{B+SB}(0)} = \frac{{\rm Tr} \big[ p_j(0) e^{-H_{\rm B +SB}/k_B T} ]}{{\rm Tr} \big[ e^{-H_{\rm B +SB}/k_B T} ]} = 0,  
    \label{avgsqm1}
\end{eqnarray} 
\begin{eqnarray}
    \left\langle \left(q_j(0)-\frac{c_j x(0)}{m_j \omega_j^2}\right)^2\right\rangle_{\rho_\mathrm{B+SB}(0)} &=& \frac{{\rm Tr} \big[  \left(q_j(0)-\frac{c_j x(0)}{m_j \omega_j^2}\right)^2 e^{-H_{\rm B +SB}/k_B T} ]}{{\rm Tr} \big[ e^{-H_{\rm B +SB}/k_B T} ]}  \nonumber \\
  &=&\frac{\hbar}{2m_j\omega_j}\coth \bigg(\frac{\hbar \omega_j}{2 k_B T}\bigg), \label{avgsqm2}
\end{eqnarray}
  \begin{eqnarray}
  \langle p^2_j(0)\rangle_{\rho_\mathrm{B+SB}(0)} &=& \frac{{\rm Tr} \big[ p_j^2(0) e^{-H_{\rm B +SB}/k_B T} ]}{{\rm Tr} \big[ e^{-H_{\rm B +SB}/k_B T} ]} = \frac{\hbar m_j \omega_j}{2} \coth \bigg(\frac{\hbar \omega_j}{2 k_B T}\bigg), \label{avgsqm3}
\end{eqnarray} and 
\begin{eqnarray}
    \left\langle \left(q_j(0)-\frac{c_j x(0)}{m_j \omega_j^2}\right) p_j(0)\right\rangle_{\rho_\mathrm{B+SB}(0)} &=& \frac{{\rm Tr} \big[  \left(q_j(0)-\frac{c_j x(0)}{m_j \omega_j^2}\right) p_j(0) e^{-H_{\rm B +SB}/k_B T} ]}{{\rm Tr} \big[ e^{-H_{\rm B +SB}/k_B T} ]} = \frac{\mathrm{i}\hbar}{2}, \nonumber \\ \label{avgsqm4}
\end{eqnarray}
\begin{eqnarray}
    \left\langle p_j(0) \left(q_j(0)-\frac{c_j x(0)}{m_j \omega_j^2}\right) \right\rangle_{\rho_\mathrm{B+SB}(0)} &=& \frac{{\rm Tr} \big[  p_j(0)\left(q_j(0)-\frac{c_j x(0)}{m_j \omega_j^2}\right) e^{-H_{\rm B +SB}/k_B T} ]}{{\rm Tr} \big[ e^{-H_{\rm B +SB}/k_B T} ]} = -\frac{\mathrm{i}\hbar}{2}, \nonumber \\ \label{avgsqm5}
\end{eqnarray}
with all the others vanishing.

\subsection{Spectral properties of the bath}\label{spectralsection}
For consistency, we should consider the collection of independent quantum oscillators in the limit $N \to \infty$, when it becomes a heat bath, and only in this limit that Eq.~(\ref{Eq.m}) becomes a bona fide Langevin equation. Now, in the limit $N\to \infty$, it is reasonable to replace any occurrence of a summation over \(j\) with an integral, allowing us to define the so-called bath spectral function, also referred to as the spectral density \(J(\omega)\), as \cite{Weiss,purisdgbook}
\begin{equation}\label{Jdef}
  J(\omega) \equiv \frac{\pi}{2} \sum_{j=1}^N \frac{c_j^2}{m_j \omega_j} \delta(\omega-\omega_j),
\end{equation}
so that Eq.~(\ref{classicalmut}) yields
\begin{equation}\label{mutJ}
  \mu(t) = \Theta(t)\frac{2}{\pi} \int_{0}^{\infty} \mathrm{d}\omega~\frac{J(\omega)}{\omega} \cos (\omega t).
\end{equation} 

A particularly simple example for $J(\omega)$ is that of Ohmic dissipation, for which the bath spectral function reads as
\begin{equation}
J(\omega) = m \gamma \omega;~~\mathrm{Ohmic~ dissipation},
\label{OhmicJ}
\end{equation}
or, equivalently, \(\mu(t)=2m\gamma \delta (t)\Theta(t)\), using $\delta(x)=(1/\pi)\int_0^\infty \mathrm{d}t~\cos(xt)$ (we are relegating to the following subsection a discussion on how Ohmic dissipation may arise in a physical situation.). This corresponds to a memory-less friction, i.e., the drag force represented by the second term on the LHS of Eq.~(\ref{Eq.m}) is instantaneous, and, on using $\int_{-\infty}^a \mathrm{d}x~\delta(x-a)=1/2$, equals $m\gamma \dot{x}(t)$.

Employing the bath spectral function~(\ref{Jdef}), we may obtain the total mass of heat-bath oscillators as
\begin{equation}\label{totm}
\sum_{j=1}^N m_j = \frac{2}{\pi} \int_0^\infty \mathrm{d}\omega~\frac{J(\omega)}{\omega^3}.
\end{equation} If the low-frequency behavior of the bath spectral function is as \(J(\omega) \sim \omega^a\), then for \(a \leq 2\), the total mass of the bath oscillators is infinite. This happens for the case of Ohmic dissipation for which one has \(a = 1\), see Eq.~(\ref{OhmicJ}). It should be noted that this divergence is an infrared divergence, and is independent of the high-frequency (ultraviolet) behavior of \(J(\omega)\).  On the other hand, let us recall that the full Hamiltonian~(\ref{Htot}) contains a term of the form \(\sum_{j=1}^N c_j^2 x^2/2m_j \omega_j^2\), which ensures that the system-bath coupling is homogeneous in space. Using the bath spectral function, one may write this term as
\begin{equation}\label{potre}
\sum_{j=1}^N \frac{c_j^2 x^2}{2m_j \omega_j^2} = \frac{x^2}{\pi}  \int_0^\infty \mathrm{d}\omega~\frac{J(\omega)}{\omega}.
\end{equation} 
Clearly, this term is divergent for Ohmic dissipation, but is finite if a suitable high-frequency cut-off is imposed. Its divergence is therefore dictated by the high-frequency behavior of \(J(\omega)\), in contrast to the divergence of the total renormalized mass of the bath oscillators that we have seen above to stem from the low-frequency behavior of \(J(\omega)\).

In addition to the above-mentioned divergences, for the quantum-mechanical case which shall be dealt-with subsequently, the Ohmic dissipation model leads to a divergence of the thermally-averaged kinetic energy of the particle \cite{ohmicdiverge}, as shall be shown later in this paper in Section (\ref{energyseriessection}). Since the high-frequency behavior appears to be problematic, leading to divergence in both the mean kinetic energy of the system and also the potential renormalization term, it is useful to introduce a cut-off frequency \(\omega_{\rm cut}\). Thus, we may consider a situation where the cut-off appears in a `Lorentzian' manner in the bath spectral function, i.e., 
\begin{equation}\label{DrudeJ}
J(\omega) = \frac{m \gamma \omega \omega_{\rm cut}^2}{\omega^2 + \omega_{\rm cut}^2}, 
\end{equation} which goes to Eq. (\ref{OhmicJ}) for \(\omega_{\rm cut} \rightarrow \infty\). A bath described by Eq. (\ref{DrudeJ}) for finite \(\omega_{\rm cut}\) shall be called a Drude bath. Notice that in either case, the total mass of the bath oscillators is infinite. Below, we mention some comments regarding the dissipation kernel and the bath-induced noise appearing in the quantum Langevin equation. 

\subsubsection{Friction kernel}
In our subsequent analysis, we will frequently encounter the Fourier transform of \(\mu(t)\), defined as (see, for instance, Refs.~\cite{quantumreg,ford,ford1988})
\begin{equation}
\widetilde{\mu}(\omega)\equiv \int_0^\infty \mathrm{d}t~\mu(t) e^{\mathrm{i}\omega t};~~\mathrm{Im}[\omega] >0,
\end{equation} where we have used the fact that \(\mu(t)\) vanishes for negative times, see Eq.~(\ref{classicalmut}). The above definition implies that \(\widetilde{\mu}(\omega)\) is analytic in the upper-half complex-$\omega$ plane.  Further, there are two more important mathematical properties that any \(\widetilde{\mu}(\omega)\) must possess. Firstly, its real part should be positive everywhere on the real-$\omega$ axis, i.e.,
\begin{equation}
\mathrm{Re} [\widetilde{\mu} (\omega)] > 0~ \forall~\omega \in (-\infty, \infty).
\end{equation} 
This requirement emerges from the second law of thermodynamics~\cite{ford1988}. The second important mathematical property that has to be satisfied by any \(\widetilde{\mu}(\omega)\) is the reality condition:
\begin{equation}
\widetilde{\mu}(\omega) = \widetilde{\mu}(-\omega)^*,
\end{equation}
where $*$ denotes complex conjugation. The above property follows from the fact that the factor \(x\) appearing in the quantum Langevin equation~(\ref{Eq.m}) is a Hermitian operator. The above-mentioned properties imply that \(\widetilde{\mu}(\omega)\) falls into a special class of functions that are known as positive-real functions. It should be clarified that positive-real functions are complex-valued functions satisfying certain special properties as mentioned above \cite{ford1988}. Furthermore, as a positive-real function in the complex-$\omega$ plane, \(\mathrm{Re}[\widetilde{\mu}(\omega)]\) has to be positive (not just analytic!) in the upper-half complex-$\omega$ plane, i.e., \(\mathrm{Re} [\widetilde{\mu}(\omega)] > 0\) for \(\mathrm{Im} [\omega] > 0\). A few more generic properties of \(\widetilde{\mu}(\omega)\) have been discussed in Ref.~\cite{ford1988}.

\subsubsection{Quantum noise}
\label{quantum-noise-sec}
In Eq.~(\ref{Eq.m}), \(f(t)\) is an operator-valued random noise with zero mean,
\begin{equation}
    \langle f(t)\rangle=0,
\end{equation}
and with the following symmetric and anti-symmetric correlation functions~\cite{purisdgbook}:
\begin{eqnarray}
\langle \lbrace f(t_1), f(t_2) \rbrace \rangle &=& \frac{2}{\pi}\int_{0}^{\infty}\mathrm{d}\omega~\hbar \omega~ \mathrm{Re} [ \widetilde{\mu} (\omega)] \coth\Big(\frac{\hbar\omega}{2k_BT}\Big)  \cos \lbrack \omega(t_1-t_2)\rbrack,  \label{symmetricnoisecorrelation1} \\
\langle \lbrack f(t_1), f(t_2) \rbrack \rangle &=& \frac{2}{\mathrm{i}\pi}\int_{0}^{\infty}\mathrm{d}\omega~\hbar \omega~\mathrm{Re}[ \widetilde{\mu} (\omega)] \sin\lbrack \omega(t_1-t_2)\rbrack.
\label{noisecommutator1}
\end{eqnarray}
 The above-mentioned properties are obtained straightforwardly, by making use of the averages (\ref{avgsqm1}), (\ref{avgsqm2}), (\ref{avgsqm3}), (\ref{avgsqm4}), and (\ref{avgsqm5}). In addition to the above properties, $f(t)$ also has the Gaussian property: the statistical average of an odd number of factors of $f(t)$ is zero, while that of an even number of factors is equal to the sum of products of pair averages with the order of the factors preserved. Equation~(\ref{noisecommutator1}) above is consistent with the fact that the noise operator commutes with itself at any given instant of time, i.e., for \(t_1 = t_2\). The classical-limit results are obtained by taking \(\hbar \rightarrow 0\) and noticing that as \(\xi \rightarrow 0\), \(\coth \xi \approx 1/\xi\).

\subsection{Weak-coupling limit}
We now briefly discuss the weak-coupling limit in the context of the quantum Langevin equation \cite{HL,ford,GSAQBM,Comm,AG_SDG,virial,agmb}, in the sense of the Born-Markov approximation. For a generic open system which is described by a total Hamiltonian \(H\), the full density operator \(\rho\) undergoes a unitary evolution as described by the Liouville–von Neumann equation: 
\begin{equation}
\mathrm{i} \hbar \frac{\mathrm{d} \rho}{\mathrm{d} t} = [H, \rho]. 
\end{equation}
In order to deduce the properties of the system alone, one requires a time-evolution equation for the reduced density operator describing the system and obtained from the full density operator \(\rho\) by performing a suitable trace over the heat-bath degrees of freedom. The latter operation, however, is not at all straightforward and often when working with master equations, one resorts to special approximations to aid in some simplification. The most commonly-encountered approximations are the so-called Born and Markov approximations, although one also finds in addition to these, the wide use of other approximations including the so-called rotating-wave and secular approximations. 

The Born approximation asserts that at the initial instant and also for subsequent times, the full density matrix is factorizable as
\begin{equation}
\rho(t) = \rho_{\rm S}(t) \otimes \rho_{\rm B},
\end{equation} where \(\rho_{\rm B}\) is the density matrix of the heat bath. This approximation rests on the `largeness' of the heat bath and the relative weakness of the coupling, which is to say the coupling of the system to the bath should not significantly alter the bath eigenstates; the bath stays at a state of equilibrium for all times and its density matrix is time independent. A further simplification is obtained under the Markov approximation, which asserts that the state of the system's density matrix is dependent only on its current state, and not on its past. In other words, the heat bath is associated with a fast dynamics and any correlations die over time really quickly. Thus, the Born-Markov approximation is essentially a weak-coupling approximation in which the system's dynamics is Markovian. These approximations, together with the rotating-wave approximation are often employed to lead to the Lindblad form of the master equations \cite{LBE}. 

Coming back to our problem of a quantum harmonic oscillator in contact with a heat bath composed of independent quantum oscillators, let us begin by noting that a consistent quantum Langevin equation is obtained only under the initial system-bath preparation where the system and the bath are taken to be coupled at the initial instant and that the density matrix is non-factorizable because of the specific form of \(\rho_{\rm B + SB}(0)\) defined in (\ref{dmclass}), meaning that there are non-trivial system-bath correlations at the initial instant. This immediately tells us that the framework of the quantum Langevin equation is devoid of the Born approximation, and therefore, the quantum Langevin equation can be used to handle strong-coupling situations as well. Moreover, the dynamics as described by the quantum Langevin equation is non-Markovian. It is then a curious question to ask whether one can implement a Born-Markov-like limit in the quantum Langevin equation of an oscillator. As it turns out, the system-bath coupling strength \(\gamma\) [Eq. (\ref{OhmicJ})] plays the role of a coarsened parameter indicating the strength of the system-bath coupling. The limit \(\gamma \rightarrow 0\) then implies most straightforwardly that \(\mu(t)\) is vanishingly small, and consequently, the full density matrix is factorizable; microscopically, this means taking the microscopic coupling constants \(\{c_j\}\) to be small, as in \(c_j << m_j \omega_j^2\). This is indeed the Born approximation for the quantum Langevin equation. It is however, impractical to set \(\gamma \rightarrow 0\) analytically in the quantum Langevin equation, as that will lead to the Heisenberg equations for an isolated quantum harmonic oscillator; the trick is to compute physically-relevant quantities and then implement in a suitable manner, the limit \(\gamma \rightarrow 0\) to recover the corresponding results for weak coupling. This will be illustrated in Sections~(\ref{thermsec}) and (\ref{fesec}) and some subtleties associated with taking such a limit is pointed out in Section~(\ref{ddsec}). It should be strongly emphasized that the limit \(\gamma \rightarrow 0\) essentially implies that \(\gamma\) is small as compared to the other relevant frequency scales of the problem but not necessarily zero, i.e., one has \(\gamma << \omega_0\); this essentially implies that there is a separation between the timescale(s) intrinsic to the system and that induced by coupling to the bath. 

We are now left with the Markov approximation which does not appear straightforward to implement indeed as the bath-induced noise is non-Markovian. To this end, we notice that in a weak-coupling scenario, the system is influenced primarily by those heat-bath excitations that resonate with the characteristic frequency of the system; this would be most transparently seen in Section~(\ref{thermsec}) [see Eq. (\ref{PKw-1})]. In other words, heat-bath oscillators with frequencies around \(\omega = \omega_0\) are the most relevant, and the system feels much less of the excitations across the wide spectrum of the heat bath. This allows us to approximate Eq. (\ref{symmetricnoisecorrelation1}) as 
\begin{equation}\label{weaknoisecorr}
\langle \lbrace f(t_1), f(t_2) \rbrace \rangle \approx 2 m \gamma \hbar \omega_0  \coth\Big(\frac{\hbar\omega_0}{2k_BT}\Big)   \delta(t_1-t_2),
\end{equation} where we picked \([ \widetilde{\mu} (\omega)]  = m \gamma\) as appropriate for Ohmic dissipation. Notice that (\ref{weaknoisecorr}) is still a quantum-mechanical result (explicit appearance of \(\hbar\)) although now, the noise correlations are approximately Markovian. Incidentally, the same noise correlations were obtained in \cite{GSAQBM} starting with a Born-Markov master equation for a quantum oscillator and by finding the corresponding quantum Langevin equations from the Fokker-Planck equation satisfied by the Wigner function. The classical limit is obtained as \(\hbar \rightarrow 0\), which gives \(\langle \lbrace f(t_1), f(t_2) \rbrace \rangle = 4 m \gamma k_B T \delta (t_1 - t_2)\). The weak-coupling limit of our problem of a quantum oscillator coupled to a heat bath shall therefore imply the Born-Markov setting described above.

\section{Quantum energy equipartition}\label{thermsec}
In this section, we formulate and study the quantum counterpart of the energy equipartition theorem for a quantum particle in a harmonic well and in contact with a quantum heat bath modeled as a collection of independent quantum oscillators. The equipartition theorem for classical systems states that when in thermal equilibrium at temperature $T$, the average of the total kinetic energy, denoted by $E_k$, is shared equally among all
the energetically-accessible degrees of freedom and which equals  $k_BT/2$ per degree of freedom. Classically, in the framework of the independent-oscillator model, in the limit in which the heat bath is an infinite collection of independent harmonic oscillators and is in equilibrium at temperature $T$, one can show that the
average of the total kinetic energy per degree of freedom of the bath and the system when both are in thermal equilibrium are equal: $\mathcal{E}_k = E_k= k_BT/2$~\cite{Boltzmann,waterson,reif}. 

In recent times, there have been research activities aimed at the articulation of the quantum analogue of the energy equipartition theorem~\cite{jarzy1,jarzy2,jarzy3,jarzy4,jarzy5,jarzy6,kaur,kaur2,kaur1,ghoshelec,kaur3,agmb}. It has been observed that unlike the classical case, the average energy of an open quantum system, modeled as a system interacting with a collection of an infinite number of independent harmonic oscillators, can
be understood as being the sum of contributions from individually-equilibrated oscillators distributed over the entire frequency spectrum of the heat bath.
The contribution to the average energy of the system from bath oscillators lying in the frequency range between $\omega$ and $\omega+\mathrm{d}\omega$ is characterized by a  probability distribution function $P(\omega)$, which is very sensitive to the microscopic details of the heat bath and the coupling between the system and the bath. In what follows, we will derive these results, by using a form of the fluctuation-dissipation theorem (FDT) due to Callen and Welton. 

\subsection{Fluctuation-dissipation theorem of Callen-Welton}
\label{Callen-Welton}
The basic set-up and steps of analysis of the FDT \`{a} la Callen and Welton are the following~\cite{Weiss,landau,fordBB,callenwelton} (see also \cite{Case,Raphael1}). Consider a quantum-mechanical system described by a generic Hamiltonian $H$, which at an initial instant $t=0$ is in thermal equilibrium at temperature $T=1/(k_B \beta)>0$. Correspondingly, the density matrix of the system is the canonical-equilibrium distribution $\rho= e^{-\beta H}/Z$, where $Z=\mathrm{Tr}[e^{-\beta H}]$ is the canonical partition function. At times $t>0$, the system is acted upon by a general time-dependent external force $f_t$ that couples to an observable $A$ of the system (e.g., $A$ could be the position of a particle). Consequently, the perturbed Hamiltonian of the system for times $t>0$ is given by $H'=H-f_tA$. Let us consider another observable $B$ of the system, which has at $t=0$ the thermal-equilibrium value $\langle B\rangle \equiv\mathrm{Tr}[B\rho]$. For times $t>0$, because of the application of the external force, the expectation value $\langle B\rangle_t \equiv \langle B\rangle(t)$ will deviate from its thermal-equilibrium value, and will be an explicit function of time. Within a linear response theory, wherein the magnitude of $f_t$ is assumed small for all times $t>0$, the deviation $\langle \Delta B(t) \rangle\equiv \langle B\rangle(t)-\langle B\rangle$ may be related to $f_t$ via the response function $\Phi_{BA}(t)$, also called the generalized susceptibility, as
\begin{equation}
    \langle \Delta B(t)\rangle=\int_{-\infty}^t \mathrm{d}s~\Phi_{BA}(t-s)f_s,
    \label{lrt1}
\end{equation}
to leading order in $f_t$. Consider the symmetrized, correlation function in the initial state of thermal equilibrium, given by 
\begin{equation}
  \Psi_{BA}(t_1-t_2)\equiv \frac{1}{2}\langle (A(t_1)B(t_2)+B(t_2)A(t_1))\rangle,  
\end{equation}
where $A(s)$ is the operator $A$ in the Heisenberg picture with respect to the unperturbed Hamiltonian $H$, and similarly for $B(s)$. That the symmetrized correlation is a function of the time difference $t_1-t_2$ follows from the fact that a state of equilibrium is invariant with respect to shift in the origin of time. The quantum fluctuation-dissipation due to Callen and Welton asserts that the Fourier transform of $\Psi_{BA}(t)$, given by $\widetilde{\Psi}_{BA}(\omega)=\int_{-\infty}^\infty \mathrm{d}t~e^{\mathrm{i}\omega t}\Psi_{BA}(t)$, is related to the Fourier transform $\widetilde{\Phi}_{BA}(\omega)$ via
\begin{equation}
\widetilde{\Psi}_{BA}(\omega)=\frac{\hbar}{2\mathrm{i}}\coth\left(\frac{\beta \hbar \omega}{2}\right)\widetilde{\Phi}_{BA}(\omega).
\end{equation}

Let us specialize to the case when $A$ and $B$ are Hermitian operators. It then follows that $\Phi_{BA}(-t)=-\Phi_{AB}(t)$, and then one can show that 
\begin{equation}
\widetilde{\Psi}_{BA}(\omega)=\frac{\hbar}{2\mathrm{i}}\coth\left(\frac{\beta \hbar \omega}{2}\right)\left[\widetilde{\Phi}_{BA}(\omega)-\widetilde{\Phi}_{BA}^\star(\omega)\right].
\end{equation}
For the case $B=A$, we get \begin{eqnarray}
\widetilde{\Psi}_{AA}(\omega)=\hbar\coth\left(\frac{\beta \hbar \omega}{2}\right)\mathrm{Im}\left[\widetilde{\Phi}_{AA}(\omega)\right].
\label{lrt2}
\end{eqnarray}
Now, it follows for the case at hand on using Eq.~(\ref{lrt1}) that $\Phi_{AA}(t)$ is real, which implies that $\mathrm{Im}[\widetilde{\Phi}_{AA}(\omega)]$ is an odd function of $\omega$. Inverse Fourier transform of Eq.~(\ref{lrt2}) then yields 
\begin{equation}
    \Psi_{AA}(t)=\frac{\hbar}{\pi} \int_0^\infty \mathrm{d}\omega~\mathrm{Im}\left[\widetilde{\Phi}_{AA}(\omega)\right]\cos(\omega t)\coth\left(\frac{\beta \hbar \omega}{2}\right).
\end{equation}

For the demonstration of quantum energy equipartition theorem, let us consider the set-up of the quantum Langevin equation considered in Section~(\ref{qbmsec}). Next, consider a weak external force $f_t$ to act on the particle and a set of weak external forces $\{f_{j,t}, j = 1, 2, \ldots , N\}$ to act on the heat-bath oscillators. Here, $f_t$ and $f_{j,t}$ are $c$-number functions of time. We take the perturbed Hamiltonian to be of the form
\begin{equation}
H'=H - xf_t-\sum_{j=1}^N f_{j,t}q_j.
\end{equation}
Using the formalism of the Callen-Welton FDT discussed above, one may then show, by first solving the equations motion of the bath oscillators and then substituting the solution in the equation of motion of the particle, that the symmetrized correlation function of the position of the particle in the equilibrium state reads as~\cite{quantumreg,fordBB}:
\begin{eqnarray}
C_{xx} (t_1-t_2) &&\equiv \frac{1}{2} \langle x (t_1) x(t_2) + x(t_2) x(t_1) \rangle \nonumber \\
&&= \frac{\hbar}{\pi} \int_0^\infty \mathrm{d}\omega~\mathrm{Im} [\alpha^{(0)} (\omega)] \cos [\omega (t_1 - t_2)] \coth \bigg(\frac{\hbar \omega}{2 k_B T}\bigg), \nonumber \\ \label{xxquantum}
\end{eqnarray}
where the generalized susceptibility $\alpha^{(0)}(\omega)\equiv \widetilde{\Phi}_{xx}(\omega)$ is given by
\begin{equation}\label{sus1111}
\alpha^{(0)}(\omega)= \frac{1}{m(\omega_0^2 - \omega^2) - \mathrm{i} \omega \widetilde{\mu}(\omega)},
\end{equation}
yielding
\begin{equation}
\mathrm{Im}[\alpha^{(0)}(\omega)]=\frac{\omega\mathrm{Re}[\widetilde{\mu}(\omega)]}{(m(\omega_0^2 - \omega^2)+\omega\mathrm{Im}[\widetilde{\mu}(\omega)])^2 +( \omega\mathrm{Re}[\widetilde{\mu}(\omega)])^2}.
\label{sus1111-1}
\end{equation}
 The result~(\ref{xxquantum}) is very general and holds even with \(\omega_0 = 0\). One may also obtain the velocity autocorrelation function in equilibrium as
\begin{eqnarray}
C_{vv} (t_1-t_2)&&\equiv\frac{1}{2} \langle v (t_1) v(t_2) + v(t_2) v(t_1) \rangle \nonumber \\&&= \frac{\hbar}{\pi}  \int_0^\infty \mathrm{d}\omega~\omega^2\mathrm{Im} [\alpha^{(0)} (\omega)]  \cos [\omega (t_1 - t_2)] \coth \bigg(\frac{\hbar \omega}{2 k_B T}\bigg). \nonumber \\\label{vvquantum}
\end{eqnarray}
In the same set-up, we may also compute such quantities as $
C_{q_jq_j} (t_1-t_2)=(1/2)\langle q_j (t_1) q_j(t_2)+ q_j(t_2) q_j(t_1) \rangle$, $
C_{\dot{q}_j\dot{q}_j} (t_1-t_2)=(1/2) \langle \dot{q}_j (t_1) \dot{q}_j(t_2)+ \dot{q}_j(t_2) \dot{q}_j(t_1) \rangle$, and $
C_{xq_j}(t_1-t_2)=(1/2) \langle x(t_1)q_j (t_2)+ q_j(t_2) x(t_1) \rangle$.

\subsection{Partitioning of energies} 
Armed with the above background, we now consider our system of interest, namely, the particle that is undergoing a dissipative dynamics following the generalized Langevin equation~(\ref{Eq.m}), and compute its average energy in equilibrium. Let us begin with the kinetic energy. Putting \(t_1 = t_2=t\) in Eq.~(\ref{vvquantum}), one obtains 
\begin{equation}
\langle v^2(t) \rangle = \frac{\hbar}{\pi} \int_0^\infty d\omega~\omega^2\mathrm{Im} [\alpha^{(0)} (\omega)]  \coth \bigg(\frac{\hbar \omega}{2 k_B T}\bigg),
\end{equation}
so that the average kinetic energy in equilibrium is
\begin{equation}\label{KE1}
E_k(T) = \frac{m\hbar}{2\pi}  \int_0^\infty \mathrm{d}\omega~\omega^2\mathrm{Im} [\alpha^{(0)} (\omega)]  \coth \bigg(\frac{\hbar \omega}{2 k_B T}\bigg).
\end{equation}
Let us note that every bath oscillator is an independent quantum oscillator, and that the average kinetic energy of a quantum oscillator in thermal equilibrium at temperature $T$ equals \
\begin{equation}
\mathcal{E}_k(\omega,T) = \frac{\hbar \omega}{4}~\coth \Big(\frac{\hbar \omega}{2 k_B T}\Big).
\label{ke-1}
\end{equation}
Defining 
\begin{equation}
P_k(\omega) \equiv \frac{2m\omega}{\pi}~\mathrm{Im}[\alpha^{(0)} (\omega)],
\label{pk}
\end{equation}
we may identify the occurrence of the factor \(\mathcal{E}_k(\omega,T) P_k (\omega)\mathrm{d}\omega\) in Eq.~(\ref{KE1}) and interpret it as the contribution to the average kinetic energy of the particle coming from those bath oscillators whose frequencies lie in the range between \(\omega\) and \(\omega + \mathrm{d}\omega\). We thus obtain
\begin{equation}\label{kineticequipartition}
E_k(T) =  \int_0^\infty \mathrm{d}\omega~\mathcal{E}_k(\omega,T) P_k (\omega),
\end{equation} 
where the factor \(P_k(\omega)\) may be interpreted as a probability density, in the sense that $P_k(\omega)\mathrm{d}\omega$ expresses the probability that oscillators with frequencies in the range between \(\omega\) and \(\omega + \mathrm{d}\omega\), and with average kinetic energy equal to $\mathcal{E}_k(\omega,T)$ contribute to the average kinetic energy of the system. It may be shown that \(P_k(\omega)\) is both positive definite and is normalized, justifying its interpretation as a bona fide probability density~(see~\ref{appB}). In recent literature, Eq.~(\ref{kineticequipartition}) has been termed as the quantum counterpart of energy equipartition theorem~\cite{jarzy1,jarzy2,jarzy3,jarzy4,jarzy5,jarzy6,kaur,kaur2,kaur1,ghoshelec,kaur3,agmb}. The justification behind this terminology can be understood from the high-temperature limit of Eq. (\ref{kineticequipartition}), for which, upon using \(\lim_{T \rightarrow \infty} \mathcal{E}_k(\omega,T) \rightarrow \frac{k_BT}{2}\), we get
\begin{equation}
\lim_{T \rightarrow \infty} E_k(T) = \frac{k_B T}{2} \int_0^\infty \mathrm{d}\omega~ P_k (\omega) = \frac{k_B T}{2},
\end{equation} yielding the familiar (classical) equipartition result, because \(P_k(\omega)\) is normalized.

Let us turn our attention to the potential energy. In Eq.~(\ref{xxquantum}), putting \(t_1 = t_2=t\), we obtain the position autocorrelation of the system in equilibrium as \begin{equation}
\langle x^2(t) \rangle = \frac{\hbar}{\pi} \int_0^\infty \mathrm{d}\omega~\mathrm{Im} [\alpha^{(0)} (\omega)]  \coth \bigg(\frac{\hbar \omega}{2 k_B T}\bigg),
\end{equation}
yielding 
the average potential energy in equilibrium as
\begin{equation}\label{PE1}
E_p(T) = \frac{m \omega_0^2 \hbar}{2\pi}  \int_0^\infty \mathrm{d}\omega~\mathrm{ Im} [\alpha^{(0)} (\omega)]  \coth \bigg(\frac{\hbar \omega}{2 k_B T}\bigg).
\end{equation}
Similar to what has been done above for the average kinetic energy, one identifies the factor
\begin{equation}
\mathcal{E}_p(\omega,T) \equiv \frac{\hbar \omega}{4}  \coth \Big(\frac{\hbar \omega}{2 k_B T}\Big)    \end{equation} as the average potential energy of a bath oscillator with frequency \(\omega\), and the factor 
\begin{equation}
    P_p (\omega) = \frac{2 m \omega_0^2}{\omega \pi}  \mathrm{Im} [\alpha^{(0)} (\omega)] \label{pp}
    \end{equation}
as the probability density for oscillators with frequency in the range between \(\omega\) and \(\omega + \mathrm{d}\omega\), and with average potential energy equal to $\mathcal{E}_p(\omega,T)$ to contribute to the average potential energy of the system. Consequently, Eq.~(\ref{PE1}) writes as
\begin{equation}\label{potentialequipartition}
E_p(T) =  \int_0^\infty \mathrm{d}\omega~\mathcal{E}_p(\omega,T) P_p (\omega),
\end{equation}
in the same vein as Eq.~(\ref{kineticequipartition}). It may be shown that similar to \(P_k(\omega)\), the quantity \(P_p(\omega)\) is a bona fide probability density function [(\ref{appB1})], and therefore Eq. (\ref{potentialequipartition}) can be interpreted along the lines of Eq. (\ref{kineticequipartition}). Notice that the energy equipartition theorem typically applies to the mean kinetic energy of a system, and not to the potential energy. The oscillator, however, is a special case where the potential energy by virtue of being quadratic in the coordinate (just like kinetic energy is quadratic in momentum) admits an analogous interpretation.

\begin{figure}
\begin{center}
\includegraphics[scale=0.88]{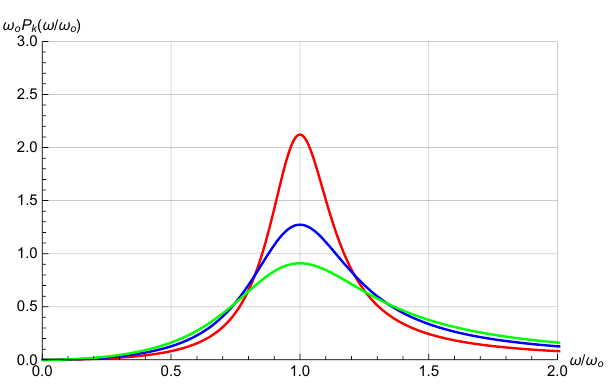}
\caption{Plot of the dimensionless factor $\omega_0P_k (\omega/\omega_0)$ characterizing the average kinetic energy, Eq.~(\ref{kineticequipartition}), as a function of $\omega/\omega_0$. Here, we have considered the case of Ohmic dissipation, $\widetilde{\mu}(\omega)=\mathrm{Re}[\widetilde{\mu}(\omega)]=m\gamma$. The parameter values are \(\gamma = 0.3 \omega_0\) (red),  \(\gamma = 0.5 \omega_0\) (blue), and \(\gamma = 0.7 \omega_0\) (green).}
\label{fig111}
\end{center}
\end{figure}

\begin{figure}
\begin{center}
\includegraphics[scale=0.88]{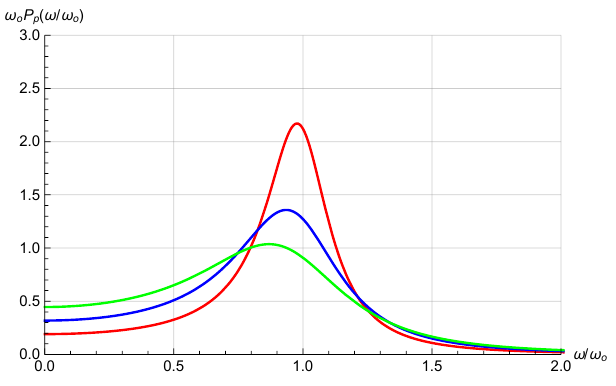}
\caption{Plot of the dimensionless factor $\omega_0P_p (\omega/\omega_0)$ characterizing the average potential energy, Eq.~(\ref{potentialequipartition}), as a function of $\omega/\omega_0$. Here, we have considered the case of Ohmic dissipation, $\widetilde{\mu}(\omega)=\mathrm{Re}[\widetilde{\mu}(\omega)]=m\gamma$. The parameter values are \(\gamma = 0.3 \omega_0\) (red),  \(\gamma = 0.5 \omega_0\) (blue), and \(\gamma = 0.7 \omega_0\) (green).}
\label{fig112}
\end{center}
\end{figure}

In Figs.~\ref{fig111} and~\ref{fig112}, we have plotted as a function of $\omega/\omega_0$ the dimensionless factors $\omega_0P_k (\omega/\omega_0)$ and  $\omega_0 P_p (\omega/\omega_0)$, respectively, for the case of Ohmic dissipation, $\mathrm{Re}[\widetilde{\mu}(\omega)]=m\gamma$, $\mathrm{Im}[\widetilde{\mu}(\omega)]=0$, and for representative values of the relevant parameters. The data are obtained by using Eqs.~(\ref{pk}),~(\ref{pp}), and~(\ref{sus1111-1}), yielding
\begin{equation}\label{pk1}
P_k (\omega) = \frac{2 \omega^2}{\pi} \frac{\gamma}{(\omega_0^2 - \omega^2)^2 + (\gamma \omega)^2} ,
\end{equation}
and
\begin{equation}\label{pp1}
P_p (\omega) = \frac{2 \omega_0^2}{\pi} \frac{\gamma}{(\omega_0^2 - \omega^2)^2 + (\gamma \omega)^2}.\end{equation}
 From the figures, we observe that as $\gamma$ gets smaller, so that the system is weakly coupled to the heat bath, the probability functions admit a sharp peak. This implies that bath oscillators whose frequencies are close to the peak frequency contribute the most to the average kinetic and potential energies of the system. Conversely, with increase of $\gamma$, the probability functions flatten out, meaning that a larger fraction of bath oscillators contribute appreciably to the average energy of the system. This feature has been analyzed in detail in Ref.~\cite{kaur3}. It may be remarked that when the particle is charged and there is a magnetic field acting on it, the number of peaks increases to two or three depending on the embedding spatial dimension of the Hamiltonian~\cite{kaur,kaur3}.

Combining the two cases discussed above, we may obtain the average of the total energy of the system in equilibrium, as
\begin{equation}\label{totalenergypartition}
E(T) = E_k(T) + E_p(T) = \int_0^\infty \mathrm{d}\omega~\mathcal{E}(\omega,T) P(\omega),
\end{equation} where \(\mathcal{E}(\omega,T) = \mathcal{E}_k(\omega,T) + \mathcal{E}_p(\omega,T)= (\hbar \omega/2)  \coth \big(\hbar \omega/(2 k_B T)\big)\) is the average of the total energy of a single bath oscillator in thermal equilibrium at temperature $T$, and we have
\begin{equation}\label{totalprob}
P (\omega) = \frac{P_k(\omega) + P_p (\omega)}{2} = \frac{m \omega}{\pi} \bigg( 1 + \frac{\omega_0^2}{\omega^2}\bigg) \mathrm{ Im} [\alpha^{(0)} (\omega)].
\end{equation}
Since \(P_k (\omega)\) and \(P_p(\omega)\) are probability density functions, their algebraic mean $P(\omega)$ is also a probability density function, and Eq.~ (\ref{totalenergypartition}) can be interpreted along the lines of Eqs.~ (\ref{kineticequipartition}) and~(\ref{potentialequipartition}). Equation~(\ref{totalenergypartition}) may be interpreted as equipartition of energy, in the sense that the average energy of the system interacting with the harmonic-oscillator heat bath is the sum of contributions from individually-equilibrated oscillators distributed over the entire frequency spectrum of the heat bath.

Finally, for the case of Ohmic dissipation, $\mathrm{Re}[\widetilde{\mu}(\omega)]=m\gamma$, $\mathrm{Im}[\widetilde{\mu}(\omega)]=0$, we remark on the weak-dissipation limit or the weak-coupling limit \(\gamma \rightarrow 0\). In this case, Eqs.~(\ref{pk1}) and~(\ref{pp1}) imply that the probability densities \(P_k(\omega)\) and \(P_p(\omega)\) individually approach a sum of delta-functions, i.e.,
\begin{equation}
P_k (\omega)= \delta(\omega - \omega_0)-\delta(\omega+\omega_0),~~P_p(\omega)= \delta(\omega - \omega_0)-\delta(\omega+\omega_0),
\label{PKw-1}
\end{equation}
and hence, Eqs.~(\ref{kineticequipartition}) and~(\ref{potentialequipartition}) yield
\begin{equation}\label{weakenergy}
E_k|_{\gamma \rightarrow 0} = E_p|_{\gamma \rightarrow 0} = \frac{\hbar \omega_0}{4} \coth \bigg(\frac{\hbar \omega_0}{2 k_B T}\bigg),
\end{equation} which are just the familiar expressions for a one-dimensional quantum oscillator in equilibrium at temperature $T$. Physically, in this limit, only the bath oscillators whose frequencies resonate with the frequency of the system contribute to the average kinetic and potential energies of the system. Thus, Eqs.~(\ref{kineticequipartition}) and~(\ref{potentialequipartition}) should be viewed as appropriate generalizations of the undergraduate textbook result given in Eq.~(\ref{weakenergy}). The reader is referred to~\cite{kaur3} for an analysis of the weak-dissipation limit in two dimensions, in presence of a magnetic field.
 

\subsection{Energy as an infinite series}\label{energyseriessection}
Quite often, the integrals appearing in Eqs.~(\ref{kineticequipartition}) and~(\ref{potentialequipartition}), or, equivalently, in Eqs.~(\ref{KE1}) and~ (\ref{PE1}), can be explicitly computed via contour integration. Utilizing the identity
\begin{equation}
z \coth z = 1+ 2\sum_{n=1}^{\infty} \frac{z^2}{(z^2)+(n\pi)^2},
\end{equation} for any complex argument \(z\), one may notice that the integrand in the integrals in Eqs.~(\ref{KE1}) and~(\ref{PE1}) have poles at \(\omega = \pm \mathrm{i} \nu_n\), where \(\nu_n = 2 \pi n k_B T/ \hbar;~n=1,2,\ldots \) are known as the bosonic Matsubara frequencies. As a result, the average kinetic and potential energies of the system turn out to be an infinite series summed over the positive integer values of \(n\). For definiteness, let us consider the case of Drude dissipation for which the bath spectral function is \(J(\omega)=m\gamma\omega/(1+(\omega/\omega_\mathrm{ cut})^2)\), with $\omega_\mathrm{ cut}$ being the upper cut-off frequency, yielding \(\widetilde{\mu}(\omega)=m\gamma\omega^2_\mathrm{ cut}/(\omega^2+\omega^2_\mathrm{ cut})+\mathrm{i}m\gamma\omega\omega_\mathrm{ cut}/(\omega^2+\omega^2_\mathrm{ cut})\). Following the aforementioned prescription, one obtains (see also \cite{ohmicdiverge})
 \begin{eqnarray}
E_k(T) &=& k_B T \Bigg[\frac{1}{2}+ \sum_{n=1}^{\infty}\frac{\omega_0^2 +\frac{\nu_n\gamma\omega_\mathrm{ cut}}{\nu_n +\omega_\mathrm{ cut}}}{\nu_n^2 +\omega_0^2 + \frac{\nu_n\gamma \omega_\mathrm{ cut}}{\nu_n +\omega_\mathrm{ cut}}} \Bigg],\label{avgke} \\
E_p(T) &=& k_BT\Bigg[\frac{1}{2} +\sum_{n=1}^{\infty}\frac{ \omega_0^2}{\nu_n^2 +\omega_0^2 + \frac{\nu_n\gamma \omega_\mathrm{ cut}}{\nu_n +\omega_\mathrm{ cut}}} \Bigg], \label{avge1}
\end{eqnarray}where $n=1,2,\cdots$. It is immediately clear that the first term in each of the above equations represents the classical result. The subsequent terms can therefore be viewed as quantum corrections. A simple inspection of Eqs.~(\ref{avgke}) and~(\ref{avge1}) tells us that the quantum corrections are always positive, implying thereby that both the average kinetic and the potential energy of the system exceed their classical counterparts. It may be verified from the above expressions that both the average kinetic and potential energies of the system are non-decreasing functions of parameters $\gamma$, $\omega_\mathrm{ cut}$, and $\omega_0$~\cite{jarzy2,kaur}. One should also note that unlike in the weak-coupling limit, the average kinetic energy is not equal to the average potential energy~\cite{virial}.

It is interesting to consider the case of Ohmic dissipation, which is obtained by considering Drude dissipation in the limit \(\omega_\mathrm{ cut}\rightarrow \infty\). Then, using Eq.~(\ref{avge1}), the average kinetic energy reads as \begin{eqnarray}\label{avgkeohmic}
E_k(T) = k_B T \Bigg[\frac{1}{2}+ \sum_{n=1}^{\infty}\frac{\omega_0^2 +\nu_n\gamma}{\nu_n^2 +\omega_0^2 + \nu_n\gamma} \Bigg],
\end{eqnarray} which is divergent~\cite{ohmicdiverge}, since the terms in the sum scale as $1/n$ for large \(n\), in contrast to Eq.~(\ref{avgke}), where the corresponding scaling is as \(\sim 1/n^2\). Such a divergence arises due to the zero-point energy of the bath oscillators, as may be verified thus: Consider just the zero-point energy contribution, i.e., \(\mathcal{E}^{(0)}_k(\omega) = \hbar \omega/4\), from the bath oscillators. This gives
\begin{equation}
E^{(0)}_k =  \int_{0}^\infty \mathrm{d}\omega~P_k (\omega) \mathcal{E}^{(0)}_k(\omega)= \frac{\hbar \gamma}{2\pi}  \int_{0}^\infty \mathrm{d}\omega ~\frac{\omega^3}{(\omega^2 - \omega_0^2)^2 + (\gamma \omega)^2},
\end{equation} which diverges! Therefore,
in situations where Ohmic dissipation is considered, it is useful to re-express Eqs.~(\ref{ke-1}),~(\ref{pk}), and~(\ref{kineticequipartition}), by omitting the zero-point energy contribution: 
\begin{equation}
E_k(T) - E^{(0)}_k = \int_0^\infty \mathrm{d}\omega~\frac{(\hbar \omega/2) P_k (\omega)}{e^{\hbar \omega/k_B T} - 1}.
\end{equation} 
This should be viewed as a suitable renormalization scheme in which
infinities have been subtracted off, leaving only the finite and physically-relevant answer. We should also remark here that the average potential energy of the particle does not suffer from a divergence when one considers the limit \(\omega_\mathrm{cut} \rightarrow \infty\) in Eq.~(\ref{avge1}).

\section{Quantum thermodynamic functions}\label{fesec}
In this section, we derive analytical expressions of various thermodynamic functions for the system described by the quantum Langevin equation~(\ref{Eq.m}), and for the most commonly discussed heat-bath model, namely, which is Ohmic. Our objective is to examine the effects of the heat bath on various thermodynamic functions, such as the internal
energy, the free energy, and the entropy of the system. 

Equation~(\ref{totalenergypartition}) gives the average energy of the system when in thermal equilibrium at temperature \(T\) while in interaction with the heat bath. There exists an alternate, inequivalent definition of energy ascribed to the system, along the lines of Refs.~\cite{lho6,FordQT,kaur2,fordBB,9,18}. This energy function \(U(T)\), which we shall refer to as internal energy of the system, can be derived as follows. 
\begin{enumerate}
\item Consider first the situation in which there is no system, but only the heat bath described by the Hamiltonian \(H_\mathrm{B}\), given by the quantum version of Eq.~(\ref{HB}). Because the bath is in thermal equilibrium at temperature $T$ for all times $t\ge 0$, it is described by the canonical-equilibrium density operator \(\rho_{H_\mathrm{B}} \propto e^{-H_\mathrm{B}/k_B T}\). We then define 
\begin{equation}
U_\mathrm{B}(T) \equiv \langle H_\mathrm{B} \rangle_{\rho_{H_\mathrm{B}}} = \frac{\mathrm{Tr} \big[H_\mathrm{B} e^{-H_\mathrm{B}/k_B T}]}{Z_\mathrm{B}}
\end{equation} as the internal energy of the free bath in its thermal-equilibrium state, with $Z_\mathrm{B}$ being the canonical partition function. Since \(H_\mathrm{B}\) models a collection of \(N\) independent quantum oscillators, we have
\begin{equation}
U_\mathrm{B}(T) = \sum_{j=1}^N \frac{\hbar \omega_j}{2} \coth \bigg(\frac{\hbar \omega_j}{2 k_B T}\bigg),
\end{equation} where the index \(j\) running from $1$ to \(N\) labels the bath oscillators. Here, we have used the well-known result that the average energy of a single quantum oscillator with frequency $\omega$ and in equilibrium at temperature $T$ equals $(\hbar \omega/2) \coth (\hbar \omega/(2 k_B T))$.

\item Next, consider a parallel situation, in which the bath is interacting with the system, so that the full Hamiltonian \(H\), given by Eq.~(\ref{Htot}), models the composite system. Given that the bath involves \(N\) degrees of freedom (with \(N\gg 1\)), \(H\) describes a system with \((N+1)\) degrees of freedom, which at thermal equilibrium is described by the canonical-equilibrium density operator  \(\rho_{H} = e^{-H/k_B T}\). Then, we may define the internal energy of this system with \((N+1)\) degrees of freedom as
\begin{equation}
\widetilde{U}(T) = \langle H \rangle_{\rho_{H}} = \frac{\mathrm{Tr} \big[ H e^{-H/k_B T}]}{Z},
\end{equation}
where $Z$ is as usual the canonical partition function. Now, we may introduce normal-mode coordinates, such that the Hamiltonian \(H\), describing a coupled \((N+1)\)-oscillator system, can be transformed into an uncoupled \((N+1)\)-oscillator system (see Ref.~\cite{lho6} for details). Consequently, \(\widetilde{U}(T)\) will be given by 
\begin{equation}
\widetilde{U}(T) = \sum_{k=0}^N \frac{\hbar \Omega_k}{2} \coth \bigg(\frac{\hbar \Omega_k}{2 k_B T}\bigg),
\end{equation} where the index \(k\) runs from 0 to \(N\) (taking into account the particle in the harmonic well), and where \(\Omega_k\)'s are the normal-mode frequencies. 
\end{enumerate}
Now that we have described what the quantities \(U_\mathrm{B}(T)\) and \(\widetilde{U}(T)\) are, we form their difference, i.e., define 
\begin{equation}\label{Udefdifference}
U(T) \equiv \widetilde{U}(T) - U_\mathrm{B}(T),
\end{equation} which can be interpreted as the internal energy of our system of particle in a harmonic well, in the sense of Ref.~\cite{lho6}.
The final expression reads (technical details of the computation can be found in \cite{lho6})
 \begin{eqnarray}
U(T) &=& \frac{1}{\pi}\int_0^{\infty}\mathrm{d}\omega~ \mathcal{E}(\omega,T)\mathrm{ Im}\Big\lbrack\frac{\mathrm d}{\mathrm{d}\omega}\ln[\alpha^{(0)}(\omega)]\Big\rbrack, \label{energyequi2}
\end{eqnarray} 
where \(\mathcal{E}(\omega,T)\) is the equilibrium average of the total energy of a bath oscillator of frequency \(\omega\):
\begin{equation}
    \mathcal{E}(\omega,T)=\frac{\hbar \omega}{2}\coth\left(\frac{\hbar \omega}{2k_BT}\right).
    \label{energyequi22}
\end{equation}
It may be pointed out that it is \(U(T)\) rather than \(E(T)\) (introduced in Section~(\ref{thermsec})) which serves as the thermodynamic energy; for the one-dimensional quantum Brownian oscillator, \(U(T)\) agrees exactly with the thermally-averaged energy obtained from the partition function of the system by evaluating Euclidean path integrals \cite{agmb}. 

For our system of interest, we may now obtain the free energy in the same spirit as above. We know that the thermally-averaged result for the free energy of a quantum oscillator of frequency \(\omega\) and in canonical equilibrium at temperature \(T\) reads
 \begin{equation}
 f(\omega,T) = k_BT \ln\bigg[2\sinh \bigg(\frac{\hbar \omega}{2 k_B T}\bigg)\bigg].
 \end{equation} With this, we define
  \begin{equation}
F_\mathrm{B}(T) = \sum_{j=1}^N f(\omega_j,T), \hspace{10mm} \widetilde{F}(T) = \sum_{k=0}^N f(\Omega_k,T),
\end{equation}  which are respectively the free energies corresponding to the situations (i) and (ii) described above (in the context of the internal energy). Consequently, the free energy for our system of a particle in a harmonic well is just their difference~\cite{lho6}: \(F_U(T) =  \widetilde{F}(T) - F_\mathrm{B}(T)\), i.e.,
\begin{equation}\label{FU11}
F_U(T) = \sum_{k=0}^N f(\Omega_k,T) - \sum_{j=1}^N f(\omega_j,T),
\end{equation} where the subscript `\(U\)' reminds the reader that it is related to the internal energy function \(U(T)\) via
\begin{equation}\label{FDef}
F_U(T) = U(T) + T \frac{\partial F_U(T)}{\partial T}.
\end{equation}  Consequently, we get
\begin{equation}\label{FT}
F_U(T) = \frac{1}{\pi} \int_0^\infty \mathrm{d}\omega~f(\omega, T) \mathrm{Im} \Bigg[ \frac{\mathrm{d}}{\mathrm{d}\omega} \ln [\alpha^{(0)}(\omega)] \Bigg].
\end{equation} 

We may alternatively discard the zero-point energy contribution in the factor \(\mathcal{E}(\omega,T)\) appearing in Eq.~(\ref{energyequi2}) and in the factor \(f(\omega,T)\) appearing in Eq.~(\ref{FT}), i.e., we can consider instead the expressions
\begin{equation}
\mathcal{E}(\omega,T) = \frac{\hbar \omega}{e^{\hbar \omega/k_B T} - 1}, \hspace{7mm} f(\omega,T) = k_B T \ln \big[ 1 - e^{-\hbar \omega/k_B T}\big]. 
\end{equation}
Then, the free energy function \(F_U(T)\) can be expressed in terms of the so-called Stieltjes \(J\)-function~\cite{FordQT,18}. 

Now, it may be shown that $\alpha^{(0)}(\omega)$ has poles on the real axis at the normal-mode frequencies $\Omega_k$ of the interacting system~(\ref{Htot}) and zeroes at the frequencies $\omega_j$ of the bath oscillators, and may be written as~\cite{lho6}
\begin{equation}
\alpha^{(0)}(\omega) = -\frac{1}{m} \frac{\prod_{j=1}^N (\omega^2 - \omega_j^2)}{\prod_{k=0}^N (\omega^2 - \Omega_k^2)}.
\end{equation}
From the well-known formula $1/(x+\mathrm{i}0^+)=\mathrm{P}(1/x)-\mathrm{i}\pi\delta(x)$, with $\mathrm{P}$ denoting the principal part, it follows that
\begin{eqnarray}\label{fnormaltheorem}
&&\frac{1}{\pi}\mathrm{ Im}\Bigg\lbrack\frac{\mathrm{d}}{\mathrm{d}\omega}\ln\alpha^{(0)}(\omega)\Bigg\rbrack\nonumber \\
 &&= \sum_{k=0}^N \big[ \delta (\omega - \Omega_k) + \delta (\omega + \Omega_k) \big]  -  \sum_{j=1}^N \big[ \delta (\omega - \omega_j) + \delta (\omega + \omega_j) \big], 
\end{eqnarray} 
inserting which into Eq.~(\ref{FT}) gives Eq.~(\ref{FU11}).

\begin{figure}
\begin{center}
\includegraphics[scale=0.88]{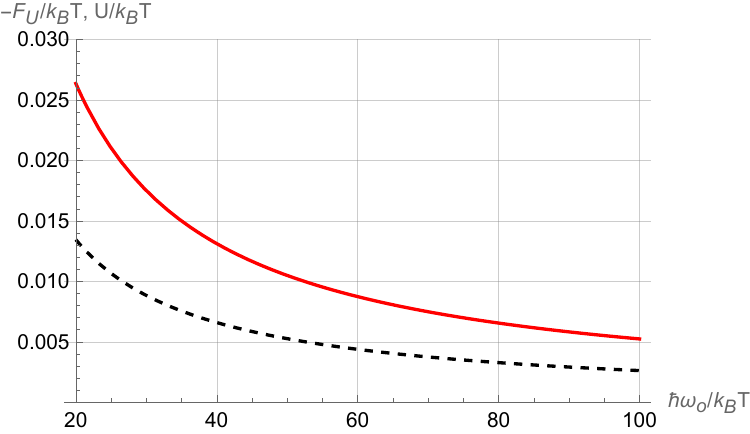}
\caption{Plot of the dimensionless factors $-F_U(T)/k_B T$ (red) and $U(T)/k_B T$ (black-dashed) as a function of the ratio \(\hbar \omega_0/k_B T\) and for \(\gamma = 0.5 \omega_0\). Only the first two terms in Eqs. (\ref{Fexpansion}) and (\ref{Uexpansion}) were taken into account in this plot.}
\label{figUF1}
\end{center}
\end{figure}

\begin{figure}
\begin{center}
\includegraphics[scale=0.88]{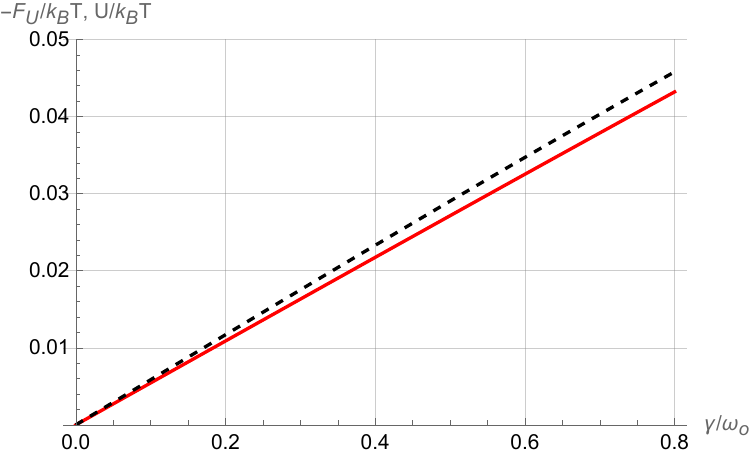}
\caption{Plot of the dimensionless factors $-F_U(T)/k_B T$ (red) and $U(T)/k_B T$ (black-dashed) as a function of the ratio \(\gamma/\omega_0 \leq 1\) and for \(\hbar \omega_0/k_B T = 10\). Only the first two terms in Eqs. (\ref{Fexpansion}) and (\ref{Uexpansion}) were taken into account in this plot.}
\label{figUF2}
\end{center}
\end{figure}

For Ohmic dissipation, one has~\cite{FordQT,kaur2}
\begin{equation}
\mathrm{ Im}\Bigg\lbrack\frac{\mathrm d}{\mathrm {d}\omega}\ln[\alpha^{(0)}(\omega)]\Bigg\rbrack = \frac{z_+}{z_+^2 + \omega^2} + \frac{z_-}{z_-^2 + \omega^2},
\end{equation} 
with \(z_{\pm} =\gamma/2 \pm \sqrt{\gamma^2/4 - \omega_0^2}\). Then, from Eq.~(\ref{FT}), after one has subtracted off the zero-point energy contribution, we get
\begin{eqnarray}
F_U(T) &=& \frac{k_B T}{\pi} \int_0^\infty \mathrm{d}\omega~\ln\Big[ 1- e^{-\hbar\omega/k_BT}\Big] \Big( \frac{z_+}{z_+^2 + \omega^2} + \frac{z_-}{z_-^2 + \omega^2} \Big) \nonumber \\
&=& -k_B T \Bigg[ J\bigg(\frac{\hbar z_+}{2 \pi k_B T}\bigg) + J\bigg(\frac{\hbar z_-}{2 \pi k_B T}\bigg) \Bigg],
\label{free-energy-Ohmic-J}
\end{eqnarray} where \(J(\zeta)\) is the Stieltjes \(J\)-function:
\begin{equation}
J(\zeta) =-\frac{1}{\pi} \int_0^\infty \mathrm{d}\xi~\ln\Big[ 1- e^{2 \pi \xi}\Big] \bigg( \frac{\zeta}{\zeta^2 + \xi^2}\bigg).
\end{equation}
Using Eq.~(\ref{free-energy-Ohmic-J}), one can obtain the behavior of various thermodynamic quantities at both low and high temperatures by utilizing expansion properties of the Stieltjes \(J\)-function. For instance, choosing the simple case of Ohmic dissipation, consider the following low-temperature (\(k_B T \ll \hbar \omega_0\)) expansions \cite{FordQT} (see also \cite{15,18}):
\begin{eqnarray}
F_U(T)  =-\bigg[ \frac{\pi (kT)^{2}\gamma }{6\hbar \omega _{0}^{2}}&+&\frac{\pi
^{3}(kT)^{4}\gamma \left( 3\omega _{0}^{2}-\gamma ^{2}\right) }{45\hbar
^{3}\omega _{0}^{6}} \nonumber  \\
&+&\frac{8\pi ^{5}(kT)^{6}\gamma \left( 5\omega
_{0}^{4}-5\gamma ^{2}\omega _{0}^{2}+\gamma ^{4}\right) }{315\hbar
^{5}\omega _{0}^{10}}+\cdots \bigg], 
\label{Fexpansion}
\end{eqnarray}  
\begin{eqnarray}
U(T) =\bigg[\frac{\pi (kT)^{2}\gamma }{6\hbar \omega _{0}^{2}}&+&\frac{\pi
^{3}(kT)^{4}\gamma \left( 3\omega _{0}^{2}-\gamma ^{2}\right) }{15\hbar
^{3}\omega _{0}^{6}} \nonumber \\
&+&\frac{8\pi ^{5}(kT)^{6}\gamma \left( 5\omega
_{0}^{4}-5\gamma ^{2}\omega _{0}^{2}+\gamma ^{4}\right) }{63\hbar
^{5}\omega _{0}^{10}}+\cdots\bigg], 
\label{Uexpansion}
\end{eqnarray} where the latter is determined from the thermodynamic definition \(U(T) = F_U(T) - T (\partial F_U(T)/\partial T)\). In Fig.~\ref{figUF1}, we have plotted \(-F_U(T)\) and \(U(T)\) in units of \(k_B T\), as a function of the ratio \(\hbar \omega_0/k_B T\) and for \(\gamma < \omega_0\). In Fig.~\ref{figUF2}, we have plotted \(-F_U(T)\) and \(U(T)\) in units of \(k_B T\), as a function of the ratio \(\gamma / \omega_0\) and for \(\hbar \omega_0/k_B T = 10\). It should be remarked that only the first two terms in Eqs. (\ref{Fexpansion}) and (\ref{Uexpansion}) were taken into account to draw these plots. The (negative) free energy and internal energy both show similar qualitative behavior, as can also be seen by inspecting Eqs.~(\ref{Fexpansion}) and~(\ref{Uexpansion}). These plots demonstrate the behavior of the energies in the low-temperature limit, which is deep into the quantum regime. The corresponding high-temperature expansions can be found in Ref.~\cite{FordQT}.


\subsection{Behavior of internal energy}
We may define the function~\cite{kaur2}
\begin{equation}\label{totalprob2}
\mathfrak{P} (\omega) \equiv  \frac{1}{\pi} \mathrm{ Im}\Big\lbrack\frac{\mathrm d}{\mathrm{d}\omega}\ln[\alpha^{(0)}(\omega)]\Big\rbrack,
\end{equation} 
which not only satisfies \(\mathfrak{P} (\omega) > 0\) for \(\omega \in [0,\infty)\), but is also normalized, i.e.,
\begin{equation}
\int_0^\infty \mathrm{d}\omega~\mathfrak{P} (\omega)= 1,
\end{equation} as can be verified from Eq.~(\ref{fnormaltheorem}). In other words, the function \(\mathfrak{P} (\omega)\) may be interpreted as a probability density function~\cite{kaur2}.
Let us then rewrite Eq.~(\ref{energyequi2}) as \begin{equation}
U(T)=\int_0^{\infty}\mathrm{d}\omega~ \mathcal{E}(\omega,T)\mathfrak{P}(\omega), \label{energyequi222}
\end{equation} 
which suggests that $\mathfrak{P}(\omega) \mathrm{d}\omega$ is the probability for oscillators with frequency in the range between $\omega$ and $\omega+\mathrm{d}\omega$, and with average equilibrium energy equal to $\mathcal{E}(\omega,T)$ to contribute to the internal energy of the system.  This suggests that the quantity \(U(T)\) satisfies a quantum counterpart of energy equipartition theorem analogous to the one satisfied by the quantity \(E(T)\) that we have studied earlier.  However, the probability density functions \(P(\omega)\) and \(\mathfrak{P}(\omega)\) differ, because \(E(T)\) and \(U(T)\) are different quantities~\cite{agmb}. It may be shown that in the weak-coupling limit, \(\mathfrak{P}(\omega)\) and \(P(\omega)\) have the same behavior, as given in Eq.~(\ref{PKw-1})~\cite{agmb}, i.e., 
\begin{equation}
P(\omega) \approx \mathfrak{P}(\omega) \approx \delta (\omega - \omega_0) + \delta (\omega + \omega_0). 
\end{equation} 
As a result, \(E(T)\) and \(U(T)\) are equal in this limit. Moreover, since \(\mathfrak{P}(\omega)\) appears in the integrand of Eq.~(\ref{FT}) for the free energy, it is found that in the weak-coupling limit, we have 
\begin{equation}
F_U(T)|_{\gamma \rightarrow 0} \approx f(\omega_0,T),
\end{equation} which is the familiar result for the free energy of a quantum oscillator at temperature \(T\) and eigenfrequency \(\omega_0\). Knowing the free energy of the system, one can obtain its entropy as \(S(T) = - \partial F_U(T)/\partial T\). 

\subsection{Entropy and the third law}\label{thirdlawsec}
The third law of thermodynamics implies that thermodynamic functions such as entropy, specific heat, isobaric coefficient of expansion,  isochoric coefficient of tension, etc., all approach zero as $T\rightarrow 0$~\cite{wilks}. Recent developments in nanotechnology have led to the development of the subject of quantum thermodynamics and low-temperature physics of small quantum
systems~\cite{myatt,capek}, and people have asked about the validity of the third law of thermodynamics in the quantum regime and what role does dissipation play. Ford and
O’Connell discussed about the third law of thermodynamics in connection with a quantum oscillator~\cite{fordphysicae2005,ocon1}. H\"anggi and Ingold have shown that finite dissipation actually helps to ensure that the third law of thermodynamics holds~\cite{hanggi,15}. Further investigations have been made in the case of dissipative cyclotron motion of a charged oscillator with different heat-bath schemes~\cite{20,momentum1,18}, and it was found that dissipative quantum systems indeed respect the third law of thermodynamics.

In order to obtain the low-temperature behavior of entropy, one can make use of the low-temperature expansion of the free energy based on an asymptotic expansion of the \(J\)-function~\cite{FordQT,20,18}. For Ohmic dissipation, this expansion gives (\ref{Fexpansion}), i.e., 
\begin{equation}
F_U(T)|_{T \rightarrow 0} = - \frac{\pi \gamma}{6 \hbar \omega_0^2} (k_BT)^2 + \mathcal{O}(T^4),
\end{equation}
implying that at low temperatures, we have on using $S(T)=-\partial F_U(T)/\partial T$ that
\begin{equation}\label{entropylinear}
   S(T)\Big|_{T \to 0} = \frac{\pi \gamma k_B^2 T}{3 \hbar \omega_0^2} + \cdots.  
\end{equation} We thus see that at low temperatures, to leading order, the entropy of our system goes to zero linearly. The same behavior is obtained even for the Drude dissipation model~\cite{FordQT}, and an identical low-temperature behavior was observed for a three-dimensional dissipative quantum oscillator placed in a magnetic field~\cite{20}. 

Let us note that from Eq.~(\ref{FT}), we have $F_U(T) = \int_0^\infty \mathrm{d}\omega~f(\omega,T) \mathfrak{P} (\omega)$. This means in the weak-coupling limit (i.e., in the limit $\gamma \rightarrow 0$), where we have $\mathfrak{P} (\omega) \approx \delta (\omega - \omega_0) + \delta(\omega + \omega_0)$, we get the entropy to be
\begin{equation}
S(T)\Big|_{\gamma \rightarrow 0} \approx - \frac{\partial}{\partial T} \int_0^\infty \mathrm{d}\omega~ f(\omega,T) \big[ \delta (\omega - \omega_0) + \delta (\omega + \omega_0) \big].
\end{equation}
Using the result \(s(\omega,T) = -\partial f(\omega,T)/\partial T\), where \(s(\omega,T)\) is the entropy of a quantum oscillator, we get
\begin{equation}\label{entropyweakcouplingexponential}
S(T)\Big|_{\gamma \rightarrow 0} \approx s(\omega_0,T). 
\end{equation}
Since the entropy of a quantum oscillator at low temperatures goes to zero
exponentially, the entropy of our system also has the same behavior in the weak-coupling limit. It should be remarked that the above-mentioned result significantly differs from that given in Eq. (\ref{entropylinear}), where we saw a linear variation of entropy with temperature, at low temperatures. This difference arises because Eq. (\ref{entropyweakcouplingexponential}) has been derived in the weak-coupling limit, while Eq. (\ref{entropylinear} has been obtained for general values of the system-bath coupling strength. Thus, finite quantum dissipation tends to convert the exponential `Einstein-like' behavior of entropy at low temperatures to a polynomial-like behavior.

\subsection{Specific heat}
Since one is discussing about thermodynamics, it is imperative to remark on the specific heat of the quantum Brownian oscillator which has been the subject of extensive investigation over the years \cite{15,hanggi,PRE79,specificheatingold,sdg2,20}. Given that \(U(T)\) is the thermodynamic energy related to the free energy \(F_U(T)\), the specific heat is defined as
\begin{equation}
C(T) = \frac{\mathrm{d}U(T)}{\mathrm{d}T} = T \frac{\mathrm{d}S(T)}{\mathrm{d}T} = -T \frac{\mathrm{d}^2F_U(T)}{\mathrm{d}T^2}.
\end{equation}
Since we have \(U(T) = \widetilde{U}(T) - U_\mathrm{B}(T) \) [Eq. (\ref{Udefdifference})], we can express the specific heat of the system as the following difference: 
\begin{equation}\label{Cdiff}
C(T) = \widetilde{C}(T) - C_\mathrm{B}(T),
\end{equation} i.e., it is the difference between the specific heats of the system + bath and the free bath \cite{specificheatingold}. In the classical limit or equivalently, at high temperatures, one simply has \(\widetilde{C} = (N+1)k_B\) and \(C_\mathrm{B} = Nk_B\), thereby giving \(C = k_B\), consistent with the Dulong-Petit law. At low temperatures, quantum effects are dominant and one has from (\ref{entropylinear}), the following low-temperature expansion in case of Ohmic dissipation: 
\begin{equation}
C(T)\Big|_{T \to 0} = \frac{\pi \gamma k_B^2 T}{3 \hbar \omega_0^2} + \cdots,
\end{equation} meaning that the specific heat goes to zero linearly as a power law. We conclude by remarking that there are various subtleties associated with the free-particle case \((\omega_0 = 0)\) which we do not address here (see, for example, Ref. \cite{15,specificheatingold}). In particular, for a free particle, the specific heat could become negative for appropriately chosen environments; this however, does not lead to thermodynamic instabilities as the quantities \(\widetilde{C}(T)\) and \(C_\mathrm{B}(T)\) are still positive with only their difference (\ref{Cdiff}) getting negative. For the harmonic oscillator which is the presently-studied case, the difference (\ref{Cdiff}) turns out to be non-negative, although under appropriate conditions one can observe a dip in the difference of specific heats
as a function of temperature \cite{specificheatingold}.

 \section{Dissipative diamagnetism}\label{ddsec}
The quantum-mechanical problem of a charged particle in presence of a magnetic field and interacting with a heat bath serves as a paradigmatic model
for the well-known Landau theory of diamagnetism.  In this section, we study Landau diamagnetism
in the context of dissipative quantum mechanics. Diamagnetism is a material property that arises from the response of a collection of non-interacting charged particles (e.g., electrons) to an applied magnetic field. The orbital magnetic moment created by the cyclotron motion of each particle exhibits a negative magnetic susceptibility, which is characteristic of diamagnetism.
The celebrated Bohr-van Leeuwen (BVL) theorem demonstrates that in the classical setting, the diamagnetic susceptibility
is identically zero as the bulk contribution to diamagnetic moment exactly cancels the contribution arising from the so-called skipping orbits of electrons (see, for example, Ref.~\cite{peirls}). However, Landau's calculation suggests that the bulk and surface contributions are different in quantum mechanics~\cite{magref1}, and hence, the cancellation of the two terms is incomplete, unlike in the classical case. This result is unusual in the sense that boundary effects, which generally do not affect thermodynamic properties, seem to play a crucial role in determining diamagnetism.

The problem of diamagnetism raises many fascinating issues concerning the inherent quantum
nature of the problem, the role played by boundaries and by dissipation, the meaning of the thermodynamic limit, and
above all, the quantum–classical crossover occasioned by environment-induced decoherence. Landau
diamagnetism provides us with a unique paradigm for discussing these relevant issues. Another issue of relevance is to be able to connect the mean orbital
magnetic moment, a thermodynamic property, with electrical resistivity, which characterizes transport
properties of materials~\cite{cursci}.

  With the motivation outlined in the preceding paragraph, we consider a particle with charge\footnote{The particle could be an electron, for which \(q = -e\), where \(e > 0\) is the magnitude of the electronic charge.} \(q\), in the presence of an external magnetic field, and in a confining harmonic potential. This system is linearly coupled to a quantum bath (modeled as a collection of independent quantum oscillators) through coordinate variables.  Thus, the Hamiltonian of the composite system is a generalization of Eq.~(\ref{Htot}), and which reads as \cite{sdg1,singh} \begin{eqnarray}
   H = \frac{(\mathbf{p} - \frac{q}{c} \mathbf{A})^2}{2m} + \frac{m \omega_0^2 \mathbf{r}^2}{2}  + \sum_{j=1}^N\bigg[\frac{\mathbf{p}_j^2}{2m_j} + \frac{1}{2}m_j \omega_j^2 \bigg( \mathbf{q}_j - \frac{c_j}{m_j \omega_j^2}\mathbf{r} \bigg)^2 \bigg],
\end{eqnarray}
where $\mathbf{p} = (p_x,p_y,p_z)$ and $\mathbf{r} = (x,y,z)$ are the momentum and position operators of the charged particle, $\mathbf{p}_j$ and $\mathbf{q}_j$ are the corresponding variables for the \(j\)-th heat-bath oscillator, and $\mathbf{A}$ is the vector potential. The usual commutation relations between coordinates and momenta, a generalization of Eq.~(\ref{canonical-commutation-0}) to three dimensions,  hold. Integrating out the
reservoir variables from the Heisenberg equations of motion of the particle, and following the set-up and the computation identical to the one employed to derive Eq.~(\ref{Eq.m}), one obtains the generalized quantum Langevin equation:
\begin{equation}\label{Eq.m11}
  m \ddot{\mathbf{r}} + \int_{-\infty}^{t} \mathrm{d}t'~\mu(t - t')~\dot{\mathbf{r}}(t')+ m \omega_0^2 \mathbf{r}-\frac{q}{c}(\dot{\mathbf{r}} \times \mathbf{B}) = \mathbf{f}(t),
\end{equation} where $\textbf{B}(\textbf{r})=\nabla \times \textbf{A}(\textbf{r})$ is the magnetic field, and \(\mu(t)\) is defined as in Eq.~(\ref{classicalmut}). We will consider the magnetic field ${\bf B}$ to be uniform
in space, with components $B_x, B_y, B_z$, and magnitude $B=\sqrt{B_x^2+B_y^2+B_z^2}$. The operator-valued random force \(\mathbf{f}(t)\) is defined in a manner analogous to the one-dimensional case (see Section~(\ref{quantum-noise-sec})), and hence, satisfies the following relations:
\begin{eqnarray}
\langle \lbrace f_{\alpha}(t_1), f_{\beta}(t_2) \rbrace \rangle &=& \delta_{\alpha\beta}\frac{2}{\pi}\int_{0}^{\infty}\mathrm{d}\omega~\hbar \omega~\mathrm{Re}[\widetilde{\mu}(\omega)]\coth\Big(\frac{\hbar\omega}{2k_BT}\Big)  \cos \lbrack \omega(t_1-t_2)\rbrack, \nonumber \\\label{symmetricnoisecorrelation} \\
\langle \lbrack f_{\alpha}(t_1), f_{\beta}(t_2) \rbrack \rangle &=& \delta_{\alpha\beta}\frac{2}{\mathrm{i}\pi}\int_{0}^{\infty}\mathrm{d}\omega~\hbar \omega~\mathrm{Re}[\widetilde{\mu}(\omega)] \sin\lbrack \omega(t_1-t_2)\rbrack,
\label{noisecommutator}
\end{eqnarray}
where the indices $\alpha$ and $\beta$ stand for the Cartesian coordinates \(x\), \(y\), and \(z\).

\subsection{Magnetic moment}\label{mmsec1}
In what follows, we shall be interested in the magnetic moment of the system and the impact of dissipation on the magnetic moment. For simplicity, we take the magnetic field to be a constant and along the \(z\) direction, i.e., \(B_x = B_y = 0\) and \(B_z = B\). Then, the particle experiences Lorentz force only in the \(x\) and \(y\) directions, while motion in the \(z\) direction is identical to that of the one-dimensional problem that was discussed in the preceding sections. For the sake of simplicity, we disregard motion in the \(z\) direction and take the particle to be confined in the \(x-y\) plane. The magnetic moment of the charged particle can be computed from the following correlation function~\cite{sdg1}:
\begin{eqnarray}
 M_z(t) = \frac{|q|}{2 c} \langle x(t) \dot{y}(t) - y(t) \dot{x}(t) \rangle  \label{mm},
\end{eqnarray} 
involving operators in the Heisenberg picture. With a few straightforward manipulations, it follows that in the stationary state (i.e., when the charged particle comes to equilibrium with the heat bath that is always in equilibrium), one has~\cite{kaur3} (see also Ref.~\cite{purisdgbook}):
\begin{equation}\label{mz}
 M_z = -\frac{|q| \hbar}{2 \pi m c}  \int_{-\infty}^\infty \mathrm{d}\omega ~\omega^2 \coth \bigg(\frac{\beta \hbar \omega}{2} \bigg) \Phi(\omega),
\end{equation} 
where the function \(\Phi(\omega)\) is defined as
\begin{eqnarray}\label{Eq.2theorem1}
 \Phi(\omega)\equiv  \frac{\mathrm{Re}[\widetilde{\gamma}(\omega)]}{\Big[\Big(\omega^2-\omega_0^2-\omega\omega_c + \omega \mathrm{Im}[\widetilde{\gamma}(\omega)]\Big)^2+(\omega \mathrm{Re}[\widetilde{\gamma}(\omega)] )^2\Big]},
\end{eqnarray} 
and we have \(\widetilde{\gamma}(\omega) = \widetilde{\mu}(\omega)/m\). Here, $\omega_c=qB/(mc)$ is 
 the usual cyclotron frequency, while $c$ is the speed of light in vacuum.

Let us now consider the case of Ohmic dissipation, for which we have \(\mathrm{Re}[\widetilde{\gamma}(\omega)] = \gamma\) and \(\mathrm{Im}[\widetilde{\gamma}(\omega)] = 0\), which yields on using Eq.~(\ref{mz}) that~\cite{sdg1,purisdgbook}
\begin{equation}\label{mz1}
 M_z = -\frac{|q| \gamma \hbar}{2 \pi m c}  \int_{-\infty}^\infty \mathrm{d}\omega  \frac{\omega^2 \coth (\hbar \omega/(2 k_B T))}{\Big[(\omega^2-\omega_0^2-\omega\omega_c)^2+(\gamma \omega )^2\Big]}.
\end{equation}
We may now choose to switch off the harmonic potential, i.e., take \(\omega_0 \rightarrow 0\) and expect that the resulting expression would correspond to the magnetic moment
of a free (quantum) Brownian particle (i.e., the charged particle in the absence of
the confining harmonic potential, which is interacting with the heat bath of quantum
harmonic oscillators). Then, in the weak-coupling limit, i.e., for \(\gamma \rightarrow 0\),  one has
\begin{eqnarray}\label{mz22}
M_z &=& -\frac{|q| \hbar}{2 m c}  \int_{-\infty}^\infty  \mathrm{d}\omega~\delta(\omega - \omega_c)  \coth \bigg(\frac{ \hbar \omega}{2 k_B T} \bigg)\nonumber \\
&=& -\frac{|q| \hbar}{2 m c} \coth \bigg(\frac{ \hbar \omega_c}{2 k_B T} \bigg),
\end{eqnarray}
which is only a part of the result one gets from Landau diamagnetism. Thus, the
limit \(\omega_0 \rightarrow 0\) applied to Eq.~(\ref{mz}) does not consistently give the correct expression for
magnetic moment consistent with Landau's final answer, which is derived by computing the partition function in a weak-coupling limit. Indeed, such an
error creeps in if one is not careful with the role of the boundary~\cite{22}. It has been
elaborated by Peierls~\cite{peirls} that it is the boundary electrons which have the so-called
‘skipping orbits’ that lead to the ‘edge currents’, and this makes a crucial contribution
to the diamagnetism.

In order to incorporate confinement effects correctly, it is important to first evaluate the integral appearing in Eq.~(\ref{mz1}) and then take the limit \(\omega_0 \rightarrow 0\)~\cite{sdg1,darwin}. Performing the integration in Eq.~(\ref{mz1}), we find~\cite{kaur3,sdg1}
\begin{eqnarray}\label{mag22}
 M_z= -\frac{2|q| k_B T}{ mc}\sum_{n=1}^{\infty} \frac{\nu_n^2\omega_c}{[\nu_n^2+\omega_0^2+\gamma\nu_n]^2+(\nu_n\omega_c)^2},
\end{eqnarray}
where $\nu_n$ are the bosonic Matsubara frequences with $n= 0, 1,2, \cdots $. The weak-coupling ($\gamma \rightarrow 0$) case can be recovered rather straightforwardly from the above result, and we obtain~\cite{PRE79,kaur3}
\begin{equation}\label{dissipationzero}
M_z|_{\gamma \rightarrow 0}=-2Bk_B T\Big(\frac{q}{mc}\Big)^2\sum_{n=1}^{\infty}\frac{\nu_n^2}{(\nu_n^2+\omega_0^2)^2+(\nu_n\omega_c)^2}.
\end{equation}
Finally, if we now switch off the harmonic trap by taking the limit $\omega_0 \rightarrow 0$ in Eq.~(\ref{dissipationzero}), we can recover the famous Landau diamagnetism result~\cite{sdg1}:
\begin{eqnarray}\label{landau}
M_z|_{\gamma, \omega_0 \rightarrow 0}= \frac{|q|\hbar}{2mc}\Bigg[\frac{2k_B T}{\hbar\omega_c}-\coth\Big(\frac{\hbar\omega_c}{2k_B T}\Big)\Bigg].
\end{eqnarray}
We emphasize that the limits \(t \rightarrow \infty\) and \(\omega_0 \rightarrow 0\) do not commute. For obtaining the above result, we have first considered the limit \(t \rightarrow \infty\) in Eq.~ (\ref{mm}) to obtain Eq.~ (\ref{dissipationzero}), followed by the limit \(\omega_0 \rightarrow 0\) in the latter equation. Reversing the order of the two limits leads to an incorrect expression. Moreover, even if we take the limit \(t \rightarrow \infty\) first to obtain Eq.~(\ref{mz}), putting \(\omega_0 \rightarrow 0\) at this stage only leads to a part of the Landau's expression [Eq.~(\ref{landau})]. This demonstrates the crucial role of confining boundaries in the context of diamagnetism. We should also remark on the importance of the ordering of the limit \(\gamma \rightarrow 0\) in the context of recovering (\ref{landau}). A quick calculation reveals that while one can invoke this limit in (\ref{mag22}) leading to (\ref{dissipationzero}) which eventually leads to the correct expression for the Landau magnetic moment in (\ref{landau}) upon taking \(\omega_0 \rightarrow 0\), taking \(\gamma \rightarrow 0\) inside the integral in (\ref{mz1}) even for a non-zero \(\omega_0\) does not consistently lead to (\ref{landau}) if one performs the integral over the delta function first and then takes \(\omega_0 \rightarrow 0\).

\begin{figure}
\begin{center}
\includegraphics[scale=0.88]{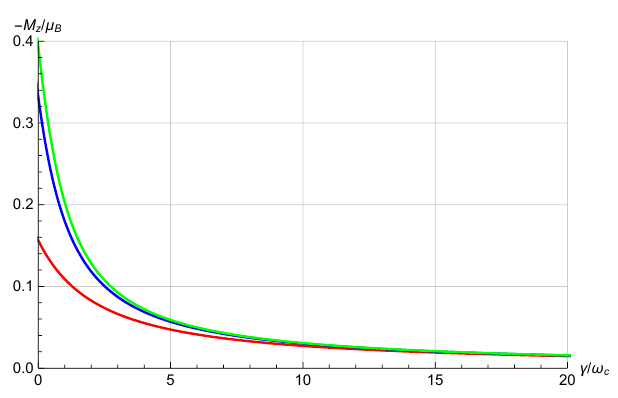}
\caption{Plot of the (negative) magnetic moment in units of Bohr magneton and as a function of the Drude resistivity \(\gamma/\omega_c\) taking \(\omega_0 = 0.5 \omega_c\), and for \(\hbar \omega_c/2k_B T = 1\) (red),  \(\hbar \omega_c/2k_B T = 3\) (blue), and \(\hbar \omega_c/2k_B T = 5\) (green).}
\label{fig222}
\end{center}
\end{figure}

Finally, one can recast Eq.~(\ref{mag22}) in the following form:
\begin{equation}\label{mag33}
M_z = -\frac{|q|\hbar}{mc}\times \frac{1}{\nu}\sum_{n=1}^{\infty}\frac{1}{\Big(\widetilde{\mu}_n+\frac{\mu_0^2}{\widetilde{\mu}_n} +\widetilde{\rho}\Big)^2+1},
\end{equation}
where the renormalized Drude resistivity $\widetilde{\rho}=\gamma/\omega_c$ is defined as the ratio of the Drude resistivity $\rho \equiv m\gamma/(pq^2)$ ($p$ is number density of the charge carrier) to the Hall resistivity $R\equiv B/(pqc)$; the renormalized Matsubara frequency is $\widetilde{\mu}_n=\nu_n/\omega_c=(\pi n)\nu$ with $\nu \equiv \hbar\omega_c/(2k_BT)$; the renormalized confining frequency is  $\mu_0=\omega_0/\omega_c$. Thus, Eq.~(\ref{mag33}) is an interesting result as it connects the orbital diamagnetic moment to the Drude resistivity~$\widetilde{\rho}$. We know that Landau diamagnetism is a result of the coherent motion of the electrons. On the other hand, as one increases Drude resistivity, the decoherence effect becomes dominant and the orbital dissipative diamagnetic moment in Eq.~(\ref{mag33}) becomes smaller~\cite{kaur3,cursci}. This ensures the validity of the Bohr-van Leeuwen theorem. As an illustration, we have plotted the magnetic moment as a function of the Drude resistivity in Fig.~\ref{fig222}, from which it is clear that increasing the level of dissipation reduces the value of the magnetic moment. This is consistent with the fact that dissipation is associated with decoherence, which acts against diamagnetism. 

\section{Momentum-momentum coupling}\label{mmcsec}
In the case of bilinear coordinate-coordinate coupling between a system and a heat bath, as discussed in the preceding sections, the coordinate of the system is
coupled to the coordinate of each bath oscillator, and we obtained a reduced
description of the particle motion in the form of a quantum
Langevin equation that is satisfied by the coordinate
operator of the particle. In this equation, coupling to the bath is described
by (i) an operator-valued random force, and (ii) a friction force
characterized by a friction kernel. As a relevant variation of the set-up, we consider in this section the complementary possibility
of coupling of a quantum system to a quantum-mechanical heat bath through the momentum variables. The resulting dissipation appearing in the reduced dynamics of the system has been termed by Leggett as anomalous dissipation~\cite{Leg}. Such a scenario has been considered previously by many authors~\cite{Leg,Andiss,Andiss0,Andiss1,momentum1,momentum}. 
To describe our set-up, we consider a gauge-invariant system-plus-reservoir model. The system comprises a charged quantum particle moving in a harmonic potential in the presence of a magnetic field. The particle is linearly coupled via the momentum variables to a quantum heat bath
consisting of independent quantum oscillators. Although there are other velocity-dependent coupling schemes (see, for example, Refs.~\cite{fordBB,9}), they can be shown to be equivalent to the standard coordinate-coordinate coupling approach via suitable transformations of the variables appearing in the total Hamiltonian~\cite{ford1988}. However, to our knowledge, this is not the case with the gauge-invariant model of momentum-momentum coupling that we consider, and therefore this situation deserves special attention. The work we review here is reported in Refs.~\cite{momentum1,momentum}.

The Hamiltonian of our gauge invariant system-plus-reservoir model has the generic form \(H = H_\mathrm{S} + H_\mathrm{B} + H_\mathrm{SB}\) that we have considered previously in the review, and is given in three spatial dimensions by \cite{momentum}
\begin{eqnarray}
    &&H=\frac{1}{2m}\Big(\textbf{p}-\frac{q}{c}{\bf A}\Big)^2+\frac{1}{2}m\omega_0^2\textbf{r}^2\nonumber \\
    &&+\sum_{j=1}^N\bigg[\frac{1}{2m_j}\Big(\textbf{p}_j-g_j\textbf{p}+\frac{g_jq}{c}\textbf{A}\Big)^2+\frac{1}{2}m_j\omega_j^2\textbf{q}_j^2\bigg], 
\label{H}
\end{eqnarray}
where $q,m, \textbf{p}, \textbf{r}$ are respectively the charge, the mass, the
momentum operator, and the coordinate operator of the particle, while
$\omega_0$ is the frequency characterizing its motion in the harmonic
potential. We take the $j$-th heat-bath oscillator to have mass $m_j$, frequency
$\omega_j$, coordinate operator $\textbf{q}_j$, and momentum operator $\textbf{p}_j$. The dimensionless parameter $g_j$ describes the coupling between the particle and the $j$-th oscillator. The vector potential $\textbf{A}(\textbf{r})$ is related to the external magnetic field $\textbf{B}(\textbf{r})$ through $\textbf{B}(\textbf{r})=\nabla \times \textbf{A}(\textbf{r})$. The various coordinate and momentum operators follow the usual commutation relations (three-dimensional generalization of Eq.~(\ref{canonical-commutation-0})). Here, as before, we will use Greek indices ($\alpha, \beta, \ldots$) to refer to the three spatial directions, while Roman indices ($i,j,k,\ldots$) will refer to the heat-bath oscillators.

Let us first demonstrate that our system of interest is gauge invariant. For this purpose, we consider the gauge transformation $\textbf{A}(\textbf{r}) \rightarrow \textbf{A}'(\textbf{r})=\textbf{A}(\textbf{r})+\nabla f(\textbf{r})$, with 
 $f(\textbf{r})$ an arbitrary function of coordinate $\textbf{r}$, and also a unitary transformation of the state vectors of the system, $|\psi(t)\rangle \rightarrow |\psi'(t)\rangle=U|\psi(t)\rangle;~U^\dagger=U^{-1}$,
 with the choice $U=\exp\Big[\frac{\mathrm{i}q}{\hbar c}f(\textbf{r})\Big]$.
Using the Hadamard formula $
e^X Y e^{-X}=Y+[X,Y]+(1/2!)[X,[X,Y]]+\ldots$, and the commutation relations between the coordinate and the momentum operators, one may easily check that $H'(\textbf{A}')=UH(\textbf{A})U^\dagger$ and hence, all physical observables remain invariant, as required.

\subsection{Quantum Langevin equation}
The Heisenberg equations of motion for the charged particle may be straightforwradly obtained from the Hamiltonian~(\ref{H}), and which reads as
\begin{eqnarray}
m_\mathrm{r}\ddot{\textbf{r}}=-m\omega_0^2\textbf{r}+\frac{q}{c}(\textbf{v}\times \textbf{B})+\frac{i\hbar
q}{2m_\mathrm{r}c}\Big(\nabla(\nabla.\textbf{A})-\nabla^2 \textbf{A}\Big)-\sum_{j=1}^N
\frac{g_j m_\mathrm{r}}{m_j}\dot{\textbf{p}}_j, \nonumber \\
\end{eqnarray}
where the renormalized mass is 
\begin{equation}
m_\mathrm{ r}=\frac{m}{1+\sum_{j=1}^{N}\frac{g_j^2m}{m_j}}.
\label{mr}
\end{equation}
On noting that $\nabla(\nabla.\textbf{A})-\nabla^2 \textbf{A}=\nabla \times (\nabla \times \textbf{A})=\nabla \times \textbf {B}=(4\pi/c)\textbf{j}$, where $\textbf{j}$ is the current producing the external magnetic field, and also the fact that in practice this current source lies outside the region where the charged particle moves, we get
\begin{equation}
m_\mathrm{r}\ddot{\textbf{r}}=-m\omega_0^2\textbf{r}+\frac{q}{c}(\textbf{v}\times \textbf
{B})-\sum_{j=1}^N \frac{g_jm_\mathrm{r}}{m_j}\dot {\textbf{p}}_j.
\label{meqn}
\end{equation}
For the heat-bath oscillators, the Heisenberg equations of motion are
\begin{eqnarray}
&&\dot{\textbf{q}}_j=\frac{1}{m_j}\Big(\textbf{p}_j-g_j\textbf{p}+\frac{g_jq}{c}\textbf{A}\Big), \\
&&\dot{\textbf{p}}_j=-m_j\omega_j^2\textbf{q}_j.
\label{pjdot}
\end{eqnarray}
Combining the above two equations and with the reasoning given in the sentence following Eq.~(\ref{mr}), we finally obtain
\begin{equation}
m_j\ddot{\textbf{q}}_j=-m_j\omega_j^2\textbf{ q}_j+g_jm\omega_0^2\textbf{r}-\frac{g_j q}{c}(\textbf{v} \times \textbf{B}).
\label{heat-bath-EOM-Bz}
\end{equation}
Then, using $\dot{\bf p}_j=-m_j\omega_j^2{\bf q}_j$ in Eq.~(\ref{meqn}) leads to 
\begin{equation}
m_\mathrm{r}\ddot {\textbf{r}}=-m\omega_0^2\textbf{r}+\frac{q}{c}(\textbf{v}\times \textbf{B})+\sum_{j=1}^N g_jm_\mathrm{r}\omega_j^2\textbf {q}_j.
\label{heqmsystem}
\end{equation}

Let us consider the magnetic field ${\bf B}$ to be uniform
in space, with components $B_x, B_y, B_z$, and magnitude $B=\sqrt{B_x^2+B_y^2+B_z^2}$. 
We now derive a quantum Langevin equation for the charged particle. To this end, we first solve the equations of motion for the bath variables.
In this case, Eq.~(\ref{heat-bath-EOM-Bz}) has the retarded solution
\begin{eqnarray}
&&\textbf{q}_j(t)=\textbf{q}^h_j(t)+\frac{g_jm\omega_0^2}{m_j\omega_j^2}\textbf{r}(t)-\frac{g_jm\omega_0^2}{m_j\omega_j^2}\textbf{r}(0)\cos(\omega_j t)\nonumber \\
&&- \frac{g_jm\omega_0^2}{m_j\omega_j^2}\int_0^t \mathrm{d}t'~\dot{{\bf r}}(t')\cos[\omega_j(t-t')]-\frac{g_jm\omega_c}{m_j\omega_j B}\Gamma \int_0^t \mathrm{d}t'~\dot{\textbf{r}}(t')\sin[\omega_j(t-t')], \nonumber \\
\label{heat-bath-solution}
\end{eqnarray}
where we have
\begin{equation}
\textbf{q}^h_j(t)\equiv\textbf{q}_j(0)\cos(\omega_jt)+\frac{\textbf{p}_j(0)}{m_j\omega_j}\sin(\omega_jt)
\end{equation}
as the contribution from the initial condition, while $\omega_c=q B/(mc)$ 
is as usual the cyclotron frequency of precessional motion of the charged particle in the magnetic field, and we have
\begin{equation}
\Gamma\equiv\begin{bmatrix}
  0 & B_z & -B_y \\
  -B_z & 0 & B_x \\
  B_y & -B_x & 0
 \end{bmatrix}.
\end{equation}

Substituting Eq.~(\ref{heat-bath-solution}) into Eq.~(\ref{heqmsystem}), we
get
\begin{equation}
m_\mathrm{r}\ddot {\textbf{r}}+\int_0^t \mathrm{d}t'~\mu(t-t')\dot{\textbf{r}}(t')+m_\mathrm
{r}\omega_0^2\textbf{r}+\mu_\mathrm{d}(t)\textbf{r}(0)-\frac{q}{c}(\textbf{v}\times \textbf{B})=\textbf{F}(t),
\label{Eqofmotion}
\end{equation}
where we have the operator-valued random force and the friction kernel given respectively by
\begin{eqnarray}
&&\textbf{F}(t)=\sum_{j=1}^N g_jm_\mathrm{r}\omega_j^2\textbf{q}^h_j(t)\Theta(t), \label{Ft} \\
&&\mu(t-t')=\mu_\mathrm{d}(t-t')+\Gamma \mu_\mathrm{od}(t-t'),
\end{eqnarray}
with $\mu_\mathrm{d}$, the diagonal part of the friction kernel $\mu$, and $\mu_\mathrm{od}$, its off-diagonal part, given by
\begin{eqnarray}
&&\mu_\mathrm{d}(t-t')\equiv\Theta(t-t')\sum_{j=1}^N \frac{g_j^2 m m_\mathrm{r}\omega_0^2}{m_j}\cos[\omega_j(t-t')], \label{mu-d} \\
&&\mu_\mathrm{od}(t-t')\equiv\Theta(t-t')\sum_{j=1}^N \frac{g_j^2 m m_\mathrm{r}\omega_j\omega_c}{m_jB}\sin[\omega_j(t-t')].\label{mu-od}
\end{eqnarray}
 
 Now, we want to demonstrate that the coordinate operator of  the charged particle
satisfies a quantum Langevin equation. To this end, let us  assume that at time $t=0$, 
there was no magnetic field, and the charged particle was held fixed at
${\bf r}(0)$, while the heat-bath oscillators 
were in canonical equilibrium at temperature $T$ with respect to the free-oscillator Hamiltonian
\begin{equation}
H_\mathrm{B}=\sum_{j=1}^N\bigg[\frac{\textbf{p}^2_j}{2m_j}+\frac{1}{2}m_j\omega_j^2\textbf{q}_j^2\bigg].
\label{HB1}
\end{equation}
The initial preparation of the system and the bath can be taken to be such that at \(t = 0\), the heat bath is in a state of thermal equilibrium with temperature \(T\) and the system is then allowed to reach a state of thermal equilibrium as time progresses. Thus, at \(t = 0\), we have for the heat-bath oscillators a canonical-equilibrium distribution at temperature $T=1/(k_B\beta)$ with respect to the Hamiltonian $H_\mathrm{B}$:
\begin{equation}\label{initialrhocoupled1}
\rho_\mathrm{B}=\frac{\exp\Big[-\beta \sum_{j=1}^N\Big[\frac{\textbf{p}^2_j(0)}{2m_j}+\frac{m_j\omega_j^2\textbf{q}_j^2(0)}{2}\Big]\Big]}{Z_\mathrm{B}},
\end{equation}
with $Z_\mathrm{B}$ being the normalizing factor or the canonical partition function.
 At a time $t \geq 0$, the particle is released, and the
magnetic field is turned on, so that subsequent evolution of the system is
controlled by Hamiltonian (\ref{H}). One may note that the preparation of the initial state is similar to the one done for the case of the coordinate-coordinate coupling in Section~(\ref{qbmsec}). Averaging with respect to the density operator~(\ref{initialrhocoupled1}), one obtains
\begin{eqnarray}
&&\langle q_{j\alpha}(0) \rangle=0, \nonumber \\
&&\langle p_{j\alpha}(0) \rangle=0, \nonumber \\
&&\langle q_{j\alpha}(0) q_{k\beta}(0) \rangle=\frac{\hbar}{2m_j\omega_j}\coth\Big(\frac{\hbar\omega_j}{2k_BT}\Big)\delta_{jk}\delta_{\alpha\beta}, \nonumber \\
&&\langle p_{j\alpha}(0) p_{k\beta}(0) \rangle=\frac{\hbar m_j\omega_j}{2}\coth\Big(\frac{\hbar\omega_j}{2k_BT}\Big)\delta_{jk}\delta_{\alpha\beta}, \nonumber \\
&&\langle q_{j\alpha}(0) p_{k\beta}(0) \rangle=-\langle p_{j\alpha}(0) q_{k\beta}(0) \rangle=\frac{1}{2}\mathrm{i}\hbar\delta_{jk}\delta_{\alpha\beta}. \nonumber \\
\label{bath-canonical-average}
\end{eqnarray}
Additionally, one has the Gaussian properties of $q_{j\alpha}(0)$ and $p_{j\alpha}(0)$, i.e., the statistical average of odd number of factors is zero, while the same for an even number of factors is equal to the sum of products of
pair averages with the order of the factors preserved. Using the results in Eq.~(\ref{bath-canonical-average}), one finds that the force operator ${\bf F}(t)$, Eq.~(\ref{Ft}), has zero mean, \(\langle \textbf{F}(t) \rangle=0\), and a symmetric correlation given by
\begin{equation}
   \frac{1}{2}\langle F_\alpha(t_1) F_\beta(t_2)+ F_\beta(t_2) F_\alpha(t_1) \rangle=\frac{\hbar\delta_{\alpha,\beta}}{2}\sum_{j=1}^N\frac{g_j^2 m_\mathrm{r}^2\omega_j^3}{m_j}\coth\Big(\frac{\hbar\omega_j}{2k_BT}\Big)\cos[\omega_j(t_1-t_2)].
\label{Ft-symmetric-correlation} 
\end{equation}
Moreover, ${\bf F}(t)$ has the Gaussian property, which follows from
the same property of the $\textbf{q}_j(0)$ and $\textbf{p}_j(0)$.
On using the canonical commutation rules between the coordinate and the momentum operators, one can further show that ${\bf F}(t)$ has the unequal-time commutator given by
\begin{equation}
\langle[F_\alpha(t_1),F_\beta(t_2)]\rangle = -\mathrm{i}\hbar\delta_{\alpha,\beta}\sum_{j=1}^N \frac{g_j^2 m_\mathrm{r}^2 \omega_j^3}{m_j}\sin[\omega_j(t_1-t_2)].
\label{Ft-unequal-time-commutator}
\end{equation}

Now, we can interpret Eq.~(\ref{Eqofmotion}) with $t \geq 0$ as a quantum Langevin equation for the particle coordinate operator with $\textbf{F}(t)$ a random force with correlation and unequal-time commutator given by Eqs.~(\ref{Ft-symmetric-correlation})
and~(\ref{Ft-unequal-time-commutator}), respectively. The friction is as usual  characterized by the friction kernel $\mu(t)$. Note that the initial
value term that depends explicitly on the
initial coordinate of the particle and the diagonal part of the
memory function can be absorbed into the
definition of the random force by defining $\textbf{f}(t) \equiv \textbf{F}(t)-\mu_\mathrm{d}(t)\textbf{r}(0)$, and then considering a situation where the initial preparation of the heat bath conforms to the distribution function $\rho^\mathrm{Shifted}_\mathrm{B}=e^{- \beta H^\mathrm{Shifted}_\mathrm{B}}/Z^\mathrm{Shifted}_\mathrm{B}$, where the `shifted' bath Hamiltonian is
\begin{equation}
H^\mathrm{Shifted}_\mathrm{B}=\sum_{j=1}^N \Bigg[\frac{\textbf{p}_j^2(0)}{2m_j}+\frac{1}{2}
m_j\omega_j^2 \bigg(\mathbf{q}_j(0)-\frac{g_j m\omega_0^2}{m_j\omega_j^2}\textbf{r}(0)\bigg)^2\Bigg].
\end{equation}
As a result, the
redefined random force $\textbf{f}(t)$ has the same statistical properties as $\textbf{F}(t).$

Finally, one may identify certain interesting features of the quantum Langevin equation~(\ref{Eqofmotion}), which are not present in the corresponding for the case of coordinate-coordinate coupling. These are: (i) the coupling with the bath renormalizes both the inertial mass as well as the harmonic potential term; (ii) the friction kernel characterizing the drag force has an  explicit magnetic-field dependence; (iii) The random force has a modified form, and its symmetric correlation and unequal-time commutator have different form than those in the case of coordinate-coordinate coupling. However, one may find a few resemblances as well. Similar to the coordinate-coordinate coupling case, the magnetic field appears in the quantum Langevin equation as a quantum-generalized classical Lorentz force term, and the random force itself has a form that is independent of the magnetic field. 

\subsection{Quantum thermodynamic functions}
In this subsection, we derive an exact formula
for the internal energy, and hence the free energy of the charged particle in
thermal equilibrium with the heat bath. Our result shows important differences in the form of
the free energy with respect to that for the coordinate-coordinate coupling. For an
illustrative heat-bath spectrum, we evaluate the free energy in the low-temperature limit,
thereby showing that the entropy of the charged particle vanishes at zero temperature, in
conformity with the third law of thermodynamics.

The free energy of the system is computed in exactly the same way as the one employed in obtaining the free energy~(\ref{FT}) for the case of coordinate-coordinate coupling. The final expression reads as 
\begin{eqnarray}
&&F_U(T,B)=\frac{1}{\pi}\int_0^\infty \mathrm{d}\omega~f(\omega,T)\mathrm{Im}\Big[\frac{\mathrm{d}}{\mathrm{d}\omega}\ln [\mathrm{
Det}~\alpha(\omega)] \nonumber \\
&&+\lambda(\omega) \Big(\frac{\omega
qB}{c}\Big)^2\Big(\frac{\mathrm{d}(G(\omega))^2}{\mathrm{d}\omega}\Big)\mathrm{Det}~\alpha(\omega)\Big],  \label{freeenergyfinal}
\end{eqnarray}
where `Det' denotes determinant, $f(\omega,T)$ is the free energy of a bath oscillator of frequency
$\omega$ (here we include the zero-point energy contribution):
\begin{equation}
f(\omega,T)=\frac{\hbar \omega}{2} + k_B T \ln \Big[1 - e^{-\hbar \omega/k_B T} \Big],  \label{fzpin}
\end{equation}
while \(G(\omega)\) is defined as
\begin{equation}
G(\omega) \equiv 1 - \sum_{j=1}^N \frac{g_j^2 m_r \omega_j^2}{m_j (\omega_j^2 - \omega^2)}.
\end{equation}
The quantity $\lambda(\omega)$ is defined as
\begin{equation}
\lambda(\omega)\equiv -m_\mathrm{r}\omega^2 + m\omega_0^2G(\omega).
\end{equation}
With these definitions, one has
\begin{eqnarray}
\alpha_{\rho \gamma}(\omega)\equiv\frac{\left[(\lambda(\omega))^2\delta_{\rho \gamma}-\left(\frac{\omega q G(\omega)}{c}\right)^2B_\rho B_\gamma-\frac{\mathrm{i} \omega q \lambda(\omega)G(\omega)}{c}\epsilon_{\rho \gamma \eta}B_\eta\right]}{\mathrm{Det}~D(\omega)},
\end{eqnarray}
and
\begin{equation}
    \mathrm{Det}~D(\omega)=\lambda(\omega)\left[(\lambda(\omega))^2-\left(\frac{\omega qBG(\omega)}{c}\right)^2\right],
\end{equation}
where $\epsilon_{\rho \sigma \eta}$ is the Levi-Civita symbol.

In order to make the role of magnetic field explicit, we write
\begin{eqnarray}
\mathrm{Det}~\alpha(\omega)&=&[\alpha^{(0)}(\omega)]^3\Big[1-\Big(\frac{\omega B
qG(\omega)}{c}\Big)^2[\alpha^{(0)}(\omega)]^2\Big]^{-1},  \label{Detalpha1}
\end{eqnarray}
where
\begin{equation}
\alpha^{(0)}(\omega)=\frac{1}{m_\mathrm{r}(\omega_0^2-\omega^2)-\mathrm{i}\omega\widetilde{\mu}_{\rm
d}(\omega)}
\end{equation}
is the susceptibility in the absence of the magnetic field and $\widetilde{\mu}_{\rm
d}(\omega)$ is the diagonal part of memory function. Using Eq.~(\ref{Detalpha1}) in Eq.~(\ref{freeenergyfinal}), we get
\begin{equation}
F_U(T,B)=F_U(T,0)+\Delta_1 F_U(T,B)+\Delta_2 F_U(T,B),
\label{freeenergydecomp}
\end{equation}
where
\begin{eqnarray}
F_U(T,0)=\frac{3}{\pi}\int_0^\infty \mathrm{d}\omega~ f(\omega,T){\rm
Im}\Big[\frac{\mathrm{d}}{\mathrm{d}\omega}\ln \big[ \alpha^{(0)}(\omega)\big]\Big]  \label{freeenergynofield}
\end{eqnarray}
is the free energy of the charged particle in the absence of the
magnetic field, which is just 3 times that given in Eq.~(\ref{FT}). The contribution from the field is contained in the two
terms $\Delta_1 F_U(T,B)$ and $\Delta_2 F_U(T,B)$, given by
\begin{eqnarray}
\hspace{-2cm}\Delta_1 F_U(T,B)=-\frac{1}{\pi}\int_0^\infty \mathrm{d}\omega~ f(\omega,T)\mathrm{Im}\Big[\frac{\mathrm{d}}{\mathrm{d}\omega}\ln \Big\{1-(G(\omega))^2\Big(\frac{\omega B
q}{c}\Big)^2[\alpha^{(0)}(\omega)]^2\Big\}\Big],\nonumber \\
\label{freeenergyfinal1}
\end{eqnarray}
and
\begin{eqnarray}
\hspace{-2cm}\Delta_2 F_U(T,B)&&=\frac{1}{\pi}\int_0^\infty \mathrm{d}\omega~ f(\omega,T){\rm
Im}\Big[[\alpha^{(0)}(\omega)]^2\Big(\frac{\omega
qB}{c}\Big)^2\nonumber \\
&&\times\Big(\frac{\mathrm{d}(G(\omega))^2}{\mathrm{d}\omega}\Big)\Big\{1-\Big(\frac{\omega B
qG(\omega)}{c}\Big)^2[\alpha^{(0)}(\omega)]^2\Big\}^{-1}\Big].
\label{freeenergyfinal2}
\end{eqnarray}

Let us now consider a concrete example of a heat bath spectrum and find the low-temperature characteristics of the free energy. We consider the case 
\begin{equation}\label{mud}
\widetilde{\mu}_\mathrm{d}(\omega)=m_{\mathrm{r}}\gamma^{1-\nu}(-\mathrm{i}\omega)^{\nu},  \hspace{4mm}    -1<\nu <1,
\end{equation}
where \(\gamma\) is a positive constant with the dimensions of frequency and $\nu$ is within the indicated range so that $\widetilde{\mu}(\omega)$ is a positive-real function \cite{momentum1}. The Ohmic, sub-Ohmic, and super-Ohmic heat baths are defined by the conditions $\nu=0$ , $-1<\nu<0$, and $0<\nu<1$, respectively. For this case, one may show that one has the leading low-temperature behavior 
\begin{equation}
F_U(T)=-3\Gamma(\nu+2)\zeta(\nu+2)\cos\Big(\frac{\nu\pi}{2}\Big)\frac{\hbar
\gamma^3}{\pi\omega_0^2}\Big(\frac{k_BT}{\hbar \gamma}\Big)^{\nu+2},
\end{equation} where \(\Gamma(z)\) is the gamma function, while \(\zeta(z)\) is the Riemann zeta function. Thus, the free energy at low temperatures is \(F_U(T) \sim T^{\nu+2}\), meaning that one has the entropy \(S = -\partial F_U(T)/\partial T \rightarrow 0\) as \(T \rightarrow 0\), in agreement with the third law. For Ohmic dissipation (\(\nu = 0\)), we have the power-law behavior \(S(T) \sim T\), which matches with that discussed in Section~(\ref{thirdlawsec}) for the case of coordinate-coordinate coupling.

\section{Fluctuation theorems}\label{FluctuationTheoremSec}
An important attribute of a quantum solid is its transport properties, as can be measured by subjecting it to a time-dependent external-force field and examining the asymptotic, stationary-state response. Because of the omnipresent field, there is constant infusion of energy into the system so that even in the stationary state, the system is not in equilibrium. Thus, we are dealing here with an `open', out-of-equilibrium system. A case in point is the celebrated Drude-Ohm geometry, in which the system concerned is an electron subjected to a monochromatic frequency-dependent electric field. When the magnitude of the electric field is weak, as in the so-called linear-response regime, the stationary-state electrical conductivity is given by the famous Kubo formula \cite{balki}:
\begin{equation}
\sigma(\omega) = e^2 \int_0^\infty \mathrm{d}t~ e^{\mathrm{i} \omega t} \langle v(0) v(t) \rangle_0,
\end{equation} where \(e\) is the electronic charge (we pick \(q = - e\), with \(e > 0\)), \(\omega\) is the frequency of the applied field \(E \cos (\omega t)\), \(v(t)\) is the velocity of the electron at time \(t\), and the angular brackets imply an autocorrelation of the velocity in the absence of the field, indicated by the subscript `0'. The classical Langevin equation for a free Brownian particle gives \cite{balki}
\begin{equation}\label{x1}
\langle v(0) v(t) \rangle_0 = \frac{k_BT}{m} e^{-t/\tau_{\rm ch}},
\end{equation}
with \(\tau_{\rm ch}\) being the relaxation time and the pre-factor is the equipartition result. Equation~(\ref{x1}) then leads to the complex conductivity as
\begin{equation}
\sigma(\omega) = \frac{e^2 k_BT}{m} \frac{\tau_{\rm ch}(1 + \mathrm{i} \omega \tau_{\rm ch})}{1 + (\omega \tau_{\rm ch})^2},
\end{equation}
where the imaginary component is responsible for dissipation.

\begin{figure}
\begin{center}
\includegraphics[scale=0.38]{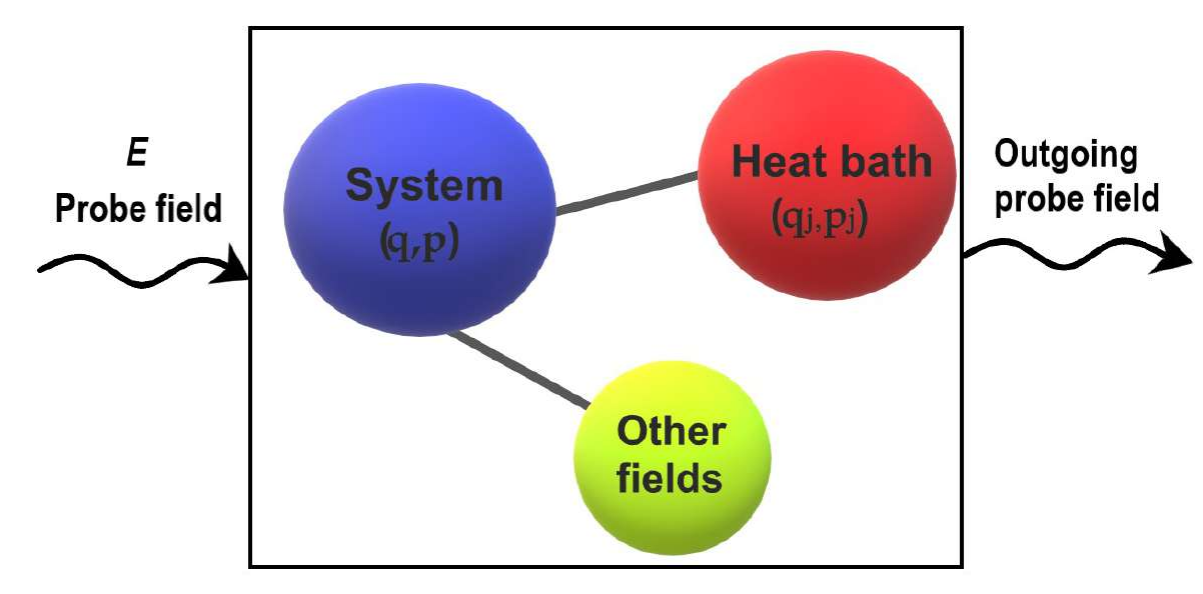}
\caption{Schematics of a typical (experimental) set-up. In the present model, the probe field is the electric field, and other fields include the magnetic field and a possible parabolic trap.}
\label{fig2}
\end{center}
\end{figure}

The question that we address now is: What happens when we go far beyond the linear-response domain, and are there general relations akin to the Kubo formula for such nonequilibrium systems? Keeping the electron paradigm in mind, the work done over a time \(t\) can be expressed as
\begin{equation}
W(t) = -e \int_0^t  \mathrm{d}t'~ E(t') v(t'),
\end{equation} where \(E(t)\) is an applied electric field (notation not be confused with energy); see Fig.~\ref{fig2} for a schematic setup. Therefore, what we propose to study in this section is the statistical property of the stochastic process \(W(t)\) within the Brownian-motion picture adopted in this review. Interestingly, there have been important developments in recent years, primarily due to the efforts of Gallavotti, Cohen, and Jarzynski~\cite{GaCoFT,JaFT,JKFT} in devising fluctuation theorems for the aforesaid systems, and we intend to provide here an overview. Once again, our model system is a particle undergoing dissipative dynamics, especially in the context of cyclotron motion - both classical (Section~(\ref{x1sec})) and quantum mechanical (Section~(\ref{x2sec})). It is well known that cyclotron motion is like a two-dimensional harmonic oscillator, and the presence of an extra dimension gives rise to possible `isometric' effects~\cite{iso1,iso2,iso3} in the context of fluctuation theorems.\\

The Gallavotti-Cohen statement of the fluctuation theorem reads \cite{GaCoFT}
\begin{equation}\label{GCFT}
\frac{p(-W)}{p(W)} = e^{-\alpha  W },
\end{equation}
where \(p(W)\) refers to the probability distribution of \(W=W(t)\). While in most thermodynamic situations, the constant \(\alpha\) does have the universal value of \(1/k_B T\), we will assess below if there is any deviation from this, due to the two-dimensional nature of cyclotron motion. For a one-dimensional process in which
\begin{equation}
W(t) = \int_0^t \mathrm{d}t'~F(t') v(t'),
\end{equation}
the theorem is easily proven as follows. Because \(v(t)\) for a Brownian particle is a Gaussian process \cite{SDBook}, so is \(W(t)\) and hence, the characteristic function \(C(h)\) defined by
\begin{equation}
C(h) = \langle e^{\mathrm{i}h W} \rangle,
\end{equation} can be decomposed as
\begin{eqnarray}
C(h) &=& e^{\mathrm{i}h \langle W \rangle} \langle e^{\mathrm{i} h \int_0^t \mathrm{d}t'~F(t') [v(t') - \langle v(t') \rangle]} \rangle \nonumber \\
&=& e^{\mathrm{i}h \langle W \rangle}  e^{ - \frac{h^2 m^2}{2} \int_0^t \mathrm{d}t'~\int_0^t \mathrm{d}t''~F(t') F(t') [\langle v(t') v(t'') \rangle - \langle v(t') \rangle \langle v(t'') \rangle]} , \label{x8}
\end{eqnarray}
where, in the last step, we have used the property of Gaussian processes as far as their cumulants are concerned \cite{SDBook}, and have written
\begin{equation}
\langle W \rangle = \int_0^t \mathrm{d}t'~F(t') \langle v(t') \rangle.
\end{equation} It should be remarked that we will be interested in the long-time limit, i.e., the limit of \(t \rightarrow \infty\), in which we will suppress the time argument; we will write \(W = W(t)\), as done above. Now, defining the variance \(\sigma^2\) by the exponent in the second term in Eq.~(\ref{x8}), we have
\begin{equation}
C(h) = e^{\mathrm{i}h \langle W \rangle} e^{\frac{-h^2 \sigma^2}{2}}.
\end{equation}
Hence, \(p(W)\), which is the Fourier transform of \(C(h)\), is also Gaussian:
\begin{equation}
p(W) = \frac{1}{\sqrt{2 \pi \sigma^2}} e^{\frac{(W - \langle W \rangle)^2}{2 \sigma^2}},
\end{equation} from which follows Eq.~(\ref{GCFT}) if we identify
\begin{equation}
\alpha = \frac{ 2 \langle W \rangle}{\sigma^2}.
\end{equation}
It remains then to enquire into the structure of \(\alpha\), as done below.

\subsection{Fluctuation theorem for classical cyclotron motion}\label{x1sec}
The Hamiltonian of an electron subject to a magnetic field \(\mathbf{B}\) along the $z$-axis is given by
\begin{equation}
H = \frac{(\mathbf{p} + \frac{e \mathbf{A}}{c})^2}{2m}.
\end{equation}
In the symmetric gauge, the vector potential \(\mathbf{A}\) has components
\begin{equation}
\mathbf{A} = \bigg(-\frac{By}{2}, \frac{Bx}{2}, 0 \bigg).
\end{equation}
The Langevin equations are given by \cite{purisdgbook}
\begin{eqnarray}
\frac{\mathrm{d}x_j}{\mathrm{d}t} = v_j, \hspace{3mm} v_j = \frac{(p_j + \frac{e A_j}{c} )}{m}, \label{maglangclas1} \\
 \frac{\mathrm{d}v_j}{\mathrm{d}t} = - \gamma v_j - \omega_c \sum_k v_k S_{kj} + \frac{f_j(t)}{m}, \label{maglangclas2}
\end{eqnarray}
where \(j =\) \(x\) or \(y\), the quantity \(\gamma\) is the damping coefficient, \(\omega_c = eB/mc\) is the cyclotron frequency, and \(S = \begin{pmatrix}
0 & 1 \\
-1  & 0 \\
\end{pmatrix}\). The \(f_j\)'s are Gaussian processes, with
\begin{equation}\label{noise2dclass11}
\langle f_j(t) \rangle =0, \hspace{5mm} \langle  f_j (t_1) f_{k}(t_2) \rangle = 2 m \gamma k_B T \delta_{j,k} \delta (t_1 - t_2).
\end{equation}

In the stationary state, the Boltzmann distribution is proportional to \(e^{\frac{-m(v_x^2 + v_y^2)}{2 k_B T}}\) and hence, we have
\begin{equation}
\langle v_x^2 \rangle = \langle v_y^2 \rangle = \frac{k_B T}{m},
\end{equation} while \(\langle v_x v_y \rangle = 0\). At this stage, we make a special assumption that the external force (an electric field in the present instance) is applied along the \(x\)-axis, as in the usual Drude geometry. As it turns out, the value of \(\alpha\) in the Gallavotti-Cohen fluctuation relation is contingent upon this stipulation and also that the damping \(\gamma\) has been chosen to be independent of the index \(j\). The mean of the work variable then is
\begin{equation}\label{x99}
\langle W(t) \rangle = \int_0^t \mathrm{d}t'~F(t') \langle v_x(t') \rangle,
\end{equation}
where \(\langle v_x(t) \rangle\) is to be obtained by solving the two-dimensional Langevin equations. The explicit solution for \(v_x(t)\) reads
\begin{equation}\label{vxsol}
v_x(t) = e^{-\gamma t} \int_0^t \mathrm{d}t'~e^{\gamma t'} \bigg\{ \Big(f_x(t') \cos [\omega_c (t-t')] - f_y(t') \sin [\omega_c (t-t')]\Big) + F(t') \cos [\omega_c (t-t')]  \bigg\}.
\end{equation}
Hence, we have
\begin{equation}
\langle v_x(t) \rangle = e^{-\gamma t} \int_0^t \mathrm{d}t' e^{\gamma t'}  F(t') \cos [\omega_c (t-t')],
\end{equation} and substituting this into Eq.~(\ref{x99}), we find
\begin{equation}\label{Wtgen}
\langle W(t) \rangle = \int_0^t \mathrm{d}t' \int_0^{t'} \mathrm{d}t''~F(t') F(t'') e^{-\gamma (t' - t'')}  \cos [\omega_c (t'-t'')] .
\end{equation}
On the other hand, the fluctuations are characterized by \(\sigma^2(t) = \langle W^2(t) \rangle - \langle W(t) \rangle^2\), where
\begin{equation}
\langle W^2(t) \rangle = \int_0^t \mathrm{d}t'~\int_0^t \mathrm{d}t^{''}~F(t') F(t'') \langle v_x(t') v_x(t'') \rangle.
\end{equation}
We first write this double integral in a time-ordered form by replacing the upper limit \(t\) of the second integral by \(t'\) and concomitantly introduce a pre-factor of 2 \cite{balki}. We then evaluate the velocity correlation from Eq.~(\ref{vxsol}) by utilizing the noise correlation from Eq.~ (\ref{noise2dclass11}), and we find with some simple manipulations, the result \cite{GSA_SDG_PRE}
\begin{equation}
 \sigma^2  = 2 k_B T \int_0^t \mathrm{d}t'~\int_0^{t'} \mathrm{d}t^{''}~F(t') F(t'') e^{-\gamma (t' - t'')}  \cos [\omega_c (t'-t'')],
\end{equation} where we have considered the long-time limit, i.e., \(t \rightarrow \infty\) and \( \sigma^2  =  \sigma^2(t) \) (in the long-time limit, we suppress the time argument). Comparing with Eq.~(\ref{Wtgen}), we get 
\begin{equation}\label{sigmatsquare}
 \sigma^2  = 2 k_B T \langle W \rangle,
\end{equation} where \(\langle W \rangle\) is the long-time limit of Eq. (\ref{Wtgen}). Thus, the parameter \(\alpha\) turns out to be
\begin{equation}\label{x25}
\alpha = \frac{1}{k_BT},
\end{equation}
validating the Gallavotti-Cohen fluctuation theorem in the present instance, even though the dissipative dynamics is in a four-dimensional phase space - two coordinates and two velocities. However, as emphasized earlier, if the friction coefficients are taken to be distinct for the motion along \(x\) and \(y\), leading to anisotropic diffusion, the parameter \(\alpha\) would deviate from its universal value in Eq.~(\ref{x25}), as is the case for isometric fluctuations \cite{iso1,iso2,iso3}.

\subsection{Fluctuation theorem for quantum cyclotron motion}\label{x2sec}
In this section, we exemplify the application of the Gallavotti-Cohen fluctuation theorem to quantum phenomena with the aid of quantum cyclotron motion. As our objective is limited to quantum Brownian motion, we begin the discussion with the corresponding quantum Langevin equations, which have the same structure as their classical counterparts in Eqs.~(\ref{maglangclas1}) and~(\ref{maglangclas2}):
\begin{eqnarray}
\frac{\mathrm{d}x_j}{\mathrm{d}t} = v_j, \hspace{3mm} v_j = \frac{(p_j + \frac{e A_j}{c} )}{m}, \label{maglangquan1} \\
 \frac{\mathrm{d}v_j}{\mathrm{d}t} = - \gamma v_j - \omega_c \sum_k v_k S_{kj} +  \frac{f_j(t)}{m}, \label{maglangquan2}
\end{eqnarray}
with however the crucial proviso that the underlying variables are non-commuting quantum operators and that \(f_j\)'s are also quantum operators with the following correlation properties (see Eq.~(\ref{noise2dclass11}), for comparison):
\begin{equation}\label{noisewccorr}
\langle f_j (t) \rangle = 0, \hspace{3mm} \langle \{ f_j(t_1) f_k(t_2) \} \rangle = 2 m \gamma \hbar \omega_c \coth \bigg(\frac{\hbar \omega_c}{2 k_B T}\bigg) \delta_{j,k} \delta(t_1-t_2).
\end{equation}

The reader is referred to Eq.~(\ref{weaknoisecorr}) for the corresponding relation for the noise correlation (in the Born-Markov approximation) for a quantum oscillator (in which the oscillator frequency \(\omega_0\) occurs in place of the cyclotron frequency \(\omega_c\)) and the detailed discussion thereafter. Evidently, this approximation is limited to the regime of large magnetic field \( \omega_c \gg \gamma\). Incidentally, this is the regime in which quantum Hall measurements are carried out wherein the electrons predominantly occupy the lowest Landau levels.

Now comes the important issue of defining the work operator in quantum mechanics, the ambiguity of which has been discussed in the literature threadbare, as it is linked with the `measurement' problem \cite{meas1,meas2,meas3,meas4}. Here we follow the path of how quantum mechanics itself had evolved from classical mechanics wherein dynamical variables were replaced by operators with commutation relations emanating from the underlying Poisson brackets and so on \cite{GSA_SDG}. Thus, assuming the Drude geometry in which the electric field \(E(t)\) (a classical field) is applied in the \(x\)-direction, we adopt the usual electrodynamics prescription of writing the work operator as
\begin{equation}
W(t) = - e \int_0^t \mathrm{d}t'~E(t') v_x(t'),
\end{equation}
where however, we have to remember the significant aspect that \(v_x(t)\) is a quantum operator which does not commute with itself at different times, and  hence its proper time ordering is essential.

Interestingly, it is only the assumption of constant damping (arising from Ohmic dissipation) that needs to be invoked in calculating the mean work. Since the Langevin equations describing the dissipative cyclotron motion, i.e., Eqs.~(\ref{maglangquan1}) and~ (\ref{maglangquan2}) have the same structure as the corresponding classical equations, i.e., Eqs.~ (\ref{maglangclas1}) and~(\ref{maglangclas2}), the solution for \(v_x(t)\) in the quantum-mechanical case is of the same form as given in Eq.~(\ref{vxsol}) with \(F(t) = - e E(t)/m\). Because even in the quantum-mechanical case, the noise averages are zero, the mean work is
\begin{equation}
\langle W(t) \rangle = \int_0^t \mathrm{d}t'~ \int_0^{t'} \mathrm{d}t''~e^{-\gamma(t' - t'')} F(t') F(t'') \cos [\omega_c (t' - t'')],
\end{equation}
which is the same result as in the classical case.

On the other hand, the calculation for the fluctuation of \(W\) will in general be much more complicated \cite{GSA_SDG} in the light of the generalized quantum Langevin equations discussed earlier, because of the non-Markovian nature of the noise correlations, notwithstanding Ohmic dissipation. However, the \textit{willy-nilly} the assumption of weak coupling in the Born-Markov sense [Eq. (\ref{noisewccorr})] leads to a rather straightforward extension of the classical result given in Eq.~(\ref{sigmatsquare}). In the long-time limit, i.e., for \(t \rightarrow \infty\), we find
\begin{equation}
 \sigma^2  = \hbar \omega_c \coth \bigg(\frac{\hbar \omega_c}{2 k_B T}\bigg) \langle W \rangle.
\end{equation}
Therefore, the fluctuation theorem retains the Gallavotti-Cohen structure of Eq.~(\ref{GCFT}), where the constant \(\alpha\) is
\begin{equation}
\alpha = \bigg[ \frac{\hbar \omega_c}{2} \coth \bigg(\frac{\hbar \omega_c}{2 k_B T}\bigg)\bigg]^{-1}.
\end{equation}
In order to delve a bit deeper into the \textit{modus operandi} of low-temperature quantum features, it is instructive to examine a specific form for the applied electric field \cite{GSA_SDG}. We assume that the latter is endowed with a frequency \(\Omega\) as also a decay parameter \(\Gamma\):
\begin{equation}
F(t) = F_0 e^{-\Gamma t} \cos (\Omega t).
\end{equation}
The choice is dictated to, by two reasons: (a) the presence of \(\Gamma\) precludes the divergence of \(\langle W \rangle\) at long times, and (b) the frequency \(\Omega\) provides a handle to tune it vis \`{a} vis the cyclotron frequency \(\omega_c\). In the long-time limit, we have
\begin{eqnarray}
\langle W \rangle &\rightarrow& \frac{F_0^2}{m \Gamma} \frac{\gamma + \Gamma}{(\gamma + \Gamma)^2 + \omega_c^2}, \hspace{16mm} \Omega = 0, \\
\langle W \rangle &\rightarrow& \frac{F_0^2}{m \Gamma} \frac{\gamma + \Gamma}{(\gamma + \Gamma)^2 + (\Omega -\omega_c)^2}, \hspace{5mm} \Omega \rightarrow \omega_c.
\end{eqnarray}
For these two limiting values of \(\Omega\), we find for the work fluctuations, the expressions
\begin{eqnarray}
 \sigma^2  &=& 2 k_B T \langle W \rangle,  \hspace{25mm} \Omega = 0, \label{x33} \\
 \sigma^2  &=& \hbar \omega_c \coth \bigg(\frac{\hbar \omega_c}{2 k_B T}\bigg) \langle W \rangle,  \hspace{6mm} \Omega \rightarrow \omega_c. \label{x34}
\end{eqnarray}
While Eq.~(\ref{x34}) merges with Eq.~(\ref{x33}) in the high-temperature limit, quantum effects prominently show up at zero temperature, where
\begin{equation}
\sigma^2  = \hbar \omega_c \langle W \rangle,  \hspace{3mm} \mathrm{for} \hspace{3mm} \Omega \rightarrow \omega_c, \hspace{2mm} T \rightarrow 0,
\end{equation}
implying that the thermal energy \(2k_BT\) gets replaced by the zero-point energy \(\hbar \omega_c\) in this extreme limit.

To summarize, the prototype of cyclotron motion in the quantum domain, which necessarily elevates the one-dimensional quantum oscillator to two dimensions, throws additional light on the validity of the fluctuation theorems, though we have not addressed here the occurrence of isometric fluctuations. Furthermore, the Born-Markov approximation, which is tantamount to assuming the system to be in weak interaction with the surrounding heat bath, seems to be able to circumvent the ticklish conundrum in defining the quantum work operator. Recall that the validity of our quantum Langevin equation(s) in the weak-coupling limit entirely hinges on a factorization approximation for the initial density matrix wherein the system begins from a nonequilibrium state while the bath is kept fixed at a definite temperature in equilibrium. In the strong-coupling situation, on the other hand, the system is never decoupled from the bath, even initially \cite{AG_SDG}.

\section{Closing remarks}\label{remarkssec}
In this review, we have presented an overview of quantum dissipative oscillators which we hope would be useful to a new entrant to the field. The study of quantum Brownian motion, inseparably linked to the quantum Langevin equation, is basic to the emerging topics of nonequilibrium statistical mechanics, dissipative quantum phenomena, and open quantum systems. Many of the themes have been widely investigated over the last four decades. As such, many of these concepts have found their way, if not in textbooks, but surely in more specialized monographs, cited here. Naturally, the bibliography in this review reflects this feature of quantum Brownian motion being a much-studied problem. However, a cursory browsing through the reference section, containing new contributions to the field, would reveal that the subject is quite alive and of active contemporary interest. Keeping this in view, the review attempts to put together a slew of topics, old and new, but related to the thermodynamic aspects, in a logical and connected manner. Unlike the problem of classical Brownian motion, new notions of non-Markov behavior, `non-white'-ness of noise, strong-coupling effects, a re-look at the third law of thermodynamics, etc., emerge in this article.

We close this article with a few remarks on coherence/decoherence, and the relevant timescales. Recall that the basic starting point in quantum mechanics is the Schr\"odinger equation for the wave function. The latter is characterized by an amplitude and a phase. The phase is the attribute that crucially sets
apart a quantum system from the corresponding classical one. It is the phase that is responsible for quantum interference phenomena, e.g., the Aharonov-Bohm effect in quantum cyclotron motion~\cite{sdg-aharonov-bohm}, not covered in this review. An important ingredient of the wave function is its time-evolution, governed by a Hamiltonian that must be hermitian, essential for inducing coherent dynamics, in the sense that there exists a definite phase relation between different states. In fact, if the system were to remain perfectly isolated, it would retain coherence indefinitely. Upon coupling the system with a heat bath, it is the system and the bath taken together that evolves in a coherent fashion; the system's coherence is shared with the environment and appears to be lost in time. This leads to the effect known as decoherence \cite{deco1,deco2}, i.e., a loss of quantum coherence, naively in the sense that the `coherent' nature of the quantum dynamics leaks into the environment. In other words, the system's dynamics as viewed in isolation (while it is coupled to the heat bath) is non-unitary and does not preserve coherence, although the evolution of the system and the bath taken together is unitary and preserves coherence. Note that in our treatment of dissipative quantum effects, the full Hamiltonian for the `open’ system reduces to just the Hamiltonian for the subsystem which indeed is hermitian, when the heat bath is decoupled. Thus the isolated-system dynamics is purely coherent that is determined by timescales intrinsic to the system. For instance, for a harmonic oscillator, that timescale is \(\omega_0^{-1}\) and for cyclotron motion,
it is \(\omega_c^{-1}\). Coupling to the heat bath introduces at least two other time scales: (i) the correlation time,
also called the relaxation time \(\tau_{\rm ch} \sim \gamma^{-1}\), and (ii) the inverse of the cut-off frequency \(\omega_{\rm cut}^{-1}\) for the bath
excitations. Notice that for Ohmic dissipation, one tacitly assumes that \(\omega_{\rm cut}\) is infinitely large. When decoherence effects are dominant, such as in a strong-coupling regime, the dynamics of the system can be described by classical mechanics rather than quantum mechanics, a feature which can be termed as a `quantum-classical crossover' (see \cite{decorev} for a recent overview). Such a quantum-classical crossover can be observed in the case of quantum cyclotron motion as discussed in Section (\ref{ddsec}), wherein in the weak-coupling regime, one has a non-zero magnetic moment (\ref{landau}) describing Landau diamagnetism. Such an effect originates from the coherent motion of charged particles (such as electrons) in the Landau levels induced by the applied magnetic field. Coupling the system to a heat bath with arbitrary coupling strength leads to the appearance decoherence in the sense that the limit \(\gamma \gg \omega_c, \omega_0\) or \(\tau_{\rm ch} \ll \omega_c^{-1}, \omega_0^{-1}\) leads to vanishing of the magnetic moment (\ref{mag22}), which is the expected classical result, as dictated by the Bohr-van Leeuwen theorem \cite{cursci} (see also \cite{kaur3}). Notice that one has not taken \(\hbar \rightarrow 0\) to achieve this; the classical-like result emerges within the full quantum-mechanical regime due to loss of coherence induced by strong dissipation. The intermediate regime between the \(\omega_c, \omega_0 \gg \gamma \approx 0\) and \(\gamma \gg  \omega_c, \omega_0\) limits is where the quantum system exhibits partial decoherence. Finally, we ought to take into account an ‘experimental timescale’ \(T_{\rm ex}\) that defines the timescales over which measurements are made. For example, if the system is driven by an oscillatory monochromatic field of frequency \(\Omega\), such as in
experiments on frequency-dependent electrical conductivity, \(T_{\rm ex}\) is essentially governed by \(\Omega^{-1}\). Naturally, for the quantum oscillator, if \(\omega_0^{-1} \ll \tau_{\rm ch}\), the system retains its coherent properties. On the other hand, if \(\omega_0^{-1} < \tau_{\rm ch}\), the
system is `weakly coherent', while for \(\omega_0^{-1} \gg \tau_{\rm ch}\), we observe complete decoherence. In the midst
of all this, \(T_{\rm ex}\) can be tuned-in to access transient (\(T_{\rm ex} < \tau_{\rm ch}\)) versus non-transient (\(T_{\rm ex} \gg \tau_{\rm ch}\)) behavior. 


\section*{Acknowledgements}
We thank the anonymous referees for many constructive comments which have led to a substantial improvement of the material presented. We are grateful to Jasleen Kaur for pointing out several typographical errors in an earlier version of the manuscript. The work of A.G. is supported by the Ministry of Education (MoE), Government of India in the form of a Prime Minister's Research Fellowship (ID: 1200454). M.B. is supported by the Department of Science and Technology (DST), Government of India under
the Core grant (Project No. CRG/2020/001768) and
MATRICS grant (Project no. MTR/2021/000566). S.D. is grateful for preliminary unpublished contributions, to Girish Agarwal, on quantum fluctuation theorems. He also thanks the Indian National Science Academy for support through their Honorary Scientist Scheme. S.G. acknowledges support from the Science and Engineering Research Board (SERB), India
under SERB-MATRICS scheme Grant No. MTR/2019/000560, and SERB-CRG scheme Grant No. CRG/2020/000596. He also thanks ICTP–Abdus Salam International Centre for Theoretical Physics, Trieste, Italy, for support under its Regular Associateship scheme.

\appendix

\section{\hspace{1.85cm} Details of potential renormalization}
\label{app11}
Here we explain why the inclusion of the term $\mathcal{V}_{\rm r}(x) = \sum_{j=1}^{N} c_j^2x^2/(2m_j\omega_j^2)$ in the Hamiltonian~(\ref{Htot}) is needed to cancel the bath-induced frequency shift. Consider the case where there is no such term, meaning that the part of the Hamiltonian summarizing the system-bath coupling [Eq. (\ref{HSB})] is now modified to
\begin{equation}
H_\mathrm{SB}=-x\sum_{j=1}^Nc_j q_j. 
\end{equation}
The Heisenberg equations now give
\begin{equation}\label{dotxqlederivation11}
m\ddot{x}(t) + m \omega_0^2 x(t) = \sum_{j=1}^N c_j q_j(t),
\end{equation}
\begin{equation}\label{qjsecondoperator11}
\ddot{q}_j(t) + \omega_j^2 q_j(t) = \frac{c_j}{m_j} x(t). 
\end{equation}
wherein (\ref{dotxqlederivation11}) differs from (\ref{dotxqlederivation}). The second equation above may be solved to give (\ref{qjsol}), which, upon substituting into (\ref{dotxqlederivation11}), we find
\begin{eqnarray}
m \ddot{x}(t) + \int_{0}^{t} \mathrm{d}t'~\mu(t - t') \dot{x}(t')+ m \omega_0^2 x(t) = f(t) + x(t) \sum_{j=1}^N \frac{c_j^2}{m_j \omega_j^2},
\end{eqnarray} where \(f(t)\) and \(\mu(t)\) have been defined in Eqs. (\ref{noiseclassicalmicro}) and (\ref{classicalmut}). We therefore observe that the frequency of the system gets shifted as
\begin{equation}
 \omega_{\rm eff}^2 =  \omega_0^2 - \frac{1}{m} \sum_{j=1}^N \frac{c_j^2}{m_j \omega_j^2},
\end{equation}  or,
\begin{equation}
\frac{m\omega_{\rm eff}^2 x^2}{2} = \bigg( \frac{m \omega_0^2}{2} -  \sum_{j=1}^N \frac{c_j^2}{2m_j \omega_j^2} \bigg)x^2. 
\end{equation} 
As observed in Sec. (\ref{spectralsection}), the frequency-correction \((\delta \omega)^2 = (1/m) \sum_{j=1}^N c_j^2/(m_j \omega_j^2) \) diverges for both Ohmic and Drude baths. This explains the inclusion of the term $\mathcal{V}_{\rm r}(x) = \sum_{j=1}^{N} c_j^2x^2/(2m_j\omega_j^2)$ in the Hamiltonian \(H_{\rm SB}\) to compensate for this bath-induced shift as explained in Sec. (\ref{qbmsec}).

\section{\hspace{1.85cm} Proof of normalization of \(P_k(\omega)\)}
\label{appB}
In this appendix, we shall explicitly prove that the function \(P_k(\omega)\) in Eq.~(\ref{pk}) is indeed positive definite and normalized. We have
\begin{equation}
P_k(\omega) = \frac{2 m \omega}{\pi} {\rm Im} [\alpha^{(0)} (\omega)],
\end{equation} or equivalently,
\begin{equation}
P_k(\omega) = \frac{2 \omega^2}{\pi} \frac{{\rm Re}[\tilde{\gamma}(\omega)]}{\Big[\Big(\omega^2-\omega_0^2+ \omega {\rm Im}[\tilde{\gamma}(\omega)]\Big)^2+(\omega {\rm Re}[\tilde{\gamma}(\omega)] )^2\Big]},
\end{equation}
which is positive definite owing to the positivity of \({\rm Re}[\tilde{\gamma}(\omega)]\).

In order to prove that $P_k(\omega)$ is normalized, let us first note the following two properties of the Laplace transform operator: 
\begin{eqnarray}
\mathcal{L} [\dot{x}(t)] &=& s \mathcal{L} [x(t)] - x(0), \\
\mathcal{L} [\ddot{x}(t)] &=& s^2 \mathcal{L} [x(t)] - s x(0) - \dot{x}(0).
\end{eqnarray}

Denoting \(\mathcal{L} [x(t)] = \hat{x}(s)\), the quantum Langevin equation [Eq.~(\ref{Eq.m})] can be expressed as (see Ref.~\cite{jarzy2} for more details)
\begin{equation}\label{langeqnlaplaced}
(ms^2  + s \hat{\mu}(s) + m \omega_0^2) \hat{x}(s) = m \dot{x}(0) + m s x(0) + \hat{f}(s).
\end{equation}
Let us denote
\begin{equation}
\mathcal{L}[Q(t)] = \hat{Q}(s) = \frac{1}{ms^2 + s \hat{\mu}(s) + m \omega_0^2}, \hspace{6mm} \mathcal{L}[R(t)] = \hat{R}(s) = \frac{ms}{ms^2 + s \hat{\mu}(s) + m \omega_0^2}.
\end{equation}
Then, Eq.~(\ref{langeqnlaplaced}) can be written as 
\begin{equation}
\hat{x}(s) = m \dot{x}(0) \hat{Q}(s) + x(0) \hat{R}(s) + \hat{Q}(s) \hat{f}(s).
\end{equation}
Equivalently, we have 
\begin{equation}
{x}(t) = p(0) Q(t) + x(0) R(t) + \int_0^t \mathrm{d}z~Q(t-z) f(z),
\end{equation} where we have \(p(0) = m \dot{x}(0)\). Similarly, one can show that the momentum operator can be expressed as
\begin{equation}\label{momentumforlaplace}
p(t) = p(0) R(t) + m x(0) \dot{R}(t) + \int_0^t \mathrm{d}z~R(t-z) f(z).
\end{equation}
Equation~(\ref{momentumforlaplace}) will allow us to determine the mean kinetic energy of the system. Now, the symmetrized momentum-momentum correlation function reads
\begin{equation}
\langle \{p(t), p(t')\} \rangle = \int_0^t \mathrm{d}t_1 \int_0^{t'} \mathrm{d}t_2~R(t - t_1) R(t' - t_2) \langle \{f(t_1), f(t_2)\} \rangle.
\end{equation} 
Using the fluctuation-dissipation theorem, this can be expressed as
\begin{equation}
\langle \{p(t), p(t')\} \rangle = \int_0^\infty \mathrm{d}\omega~ \widetilde{C}_{ff}(\omega) \int_0^t \mathrm{d}t_1 \int_0^{t'} \mathrm{d}t_2~R(t - t_1) R(t' - t_2) \cos [\omega(t_1-t_2)],
\end{equation} where \(\widetilde{C}_{ff}(\omega) = (\hbar \omega/2) \coth \big(\hbar \omega/(2 k_B T)\big) \mathrm{Re}[\widetilde{\mu}(\omega)\)] is the cosine transform of the noise autocorrelation function. Next, we put \(t = t'\), giving us
\begin{equation}
\langle p^2(t) \rangle = \int_0^\infty \mathrm{d}\omega~\widetilde{C}_{ff}(\omega) \int_0^t \mathrm{d}\tau \int_0^{t} \mathrm{d}u~R(\tau) R(u) \cos [\omega(\tau-u)],
\end{equation} where we have put \(\tau = t - t_1\) and \(u = t - t_2\). For the stationary state, we now take the limit \(t \rightarrow \infty\), which gives 
\begin{equation}\label{Ektgoestoinfty}
E_k = \lim_{t \rightarrow \infty} \frac{\langle p(t)^2 \rangle}{2m} = \frac{1}{2m} \int_0^\infty \mathrm{d}\omega~\widetilde{C}_{ff}(\omega) I(\omega),
\end{equation} with 
\begin{eqnarray}
I(\omega) &=& \int_0^\infty \mathrm{d}\tau \int_0^{\infty} \mathrm{d}u~R(\tau) R(u) \cos [\omega(\tau-u)] \nonumber \\
&=& \frac{1}{2} \int_0^\infty \mathrm{d}\tau~R(\tau) e^{\mathrm{i} \omega \tau} \int_0^\infty \mathrm{d}u~R(u) e^{- \mathrm{i} \omega u} \nonumber \\
&&+ \frac{1}{2} \int_0^\infty \mathrm{d}\tau~ R(\tau) e^{-\mathrm{i} \omega \tau} \int_0^\infty \mathrm{d}u~R(u) e^{ \mathrm{i} \omega u} \nonumber \\
&=& \hat{R}(\mathrm{i} \omega) \hat{R} (-\mathrm{i}\omega). 
\end{eqnarray}
Substituting this into Eq.~(\ref{Ektgoestoinfty}), we find
\begin{equation}
E_k = \frac{1}{2m} \int_0^\infty \mathrm{d}\omega~ \widetilde{C}_{ff}(\omega)  \hat{R}(\mathrm{i} \omega) \hat{R} (-\mathrm{i}\omega). 
\end{equation}
This means that we can identify
\begin{eqnarray}
P_k(\omega) &=& \frac{1}{m} {\rm Re} [\widetilde{\mu}(\omega)] \hat{R}(\mathrm{i} \omega) \hat{R} (-\mathrm{i}\omega) \nonumber \\
&=& [\hat{R}(\mathrm{i} \omega) + \hat{R} (-\mathrm{i}\omega)] \equiv R_F (\omega).
\end{eqnarray}
It turns out that \(R_F(\omega)\) is the cosine transform of \(R(t)\), i.e., 
\begin{equation}
R_F(\omega) = \frac{2}{\pi} \int_0^\infty \mathrm{d}t~R(t) \cos (\omega t),
\end{equation} implying 
\begin{equation}
R(t) = \int_0^\infty \mathrm{d}\omega~R_F(\omega) \cos (\omega t).
\end{equation}
Putting \(t = 0\), we find 
\begin{equation}
R(0) = \int_0^\infty \mathrm{d}\omega~R_F(\omega). 
\end{equation}
We now use the initial-value theorem of the Laplace transform to give 
\begin{equation}
R(0) = \lim_{s \rightarrow \infty}  s \hat{R}(s) = 1,
\end{equation} thereby giving
\begin{equation}
 \int_0^\infty \mathrm{d}\omega~R_F(\omega)=  \int_0^\infty \mathrm{d}\omega~P_k(\omega)= 1. 
\end{equation}

\section{\hspace{1.85cm} Proof of normalization of \(P_p(\omega)\)}
\label{appB1}
In this appendix, we will prove that the function \(P_p(\omega)\) in Eq.~(\ref{pp}) is positive definite and normalized. We have
\begin{equation}
P_p(\omega) = \frac{2 m \omega_0^2}{\pi \omega} {\rm Im} [\alpha^{(0)} (\omega)],
\end{equation} or equivalently,
\begin{equation}
P_p(\omega) = \frac{2 \omega_0^2}{\pi} \frac{{\rm Re}[\tilde{\gamma}(\omega)]}{\Big[\Big(\omega^2-\omega_0^2+ \omega {\rm Im}[\tilde{\gamma}(\omega)]\Big)^2+(\omega {\rm Re}[\tilde{\gamma}(\omega)] )^2\Big]},
\end{equation}
which is positive definite owing to the positivity of \({\rm Re}[\tilde{\gamma}(\omega)]\). In order to verify that \(P_p(\omega)\) is normalized, consider the integral
\begin{equation}\label{normalcondpp}
\int_0^\infty \mathrm{d}\omega~P_p (\omega)= \frac{2 m \omega_0^2}{\pi} \int_0^\infty \mathrm{d}\omega~\frac{{\rm Im} [\alpha^{(0)} (\omega)] }{\omega}.
\end{equation}
Next, we recall the following relationship from \cite{landau}:
\begin{equation}
\alpha^{(0)} (\mathrm{i} \omega) = \frac{2}{\pi} \int_0^\infty \mathrm{d}s~\frac{{\rm Im} [\alpha^{(0)} (s)] }{\omega^2 + s^2} s,
\end{equation} which means putting \(\omega =0\) gives us
\begin{equation}\label{normalpp2}
\alpha^{(0)} (0) = \frac{2}{\pi} \int_0^\infty \mathrm{d}\omega~\frac{{\rm Im} [\alpha^{(0)} (s)] }{s}.
\end{equation} Therefore, if we combine Eqs.~(\ref{normalcondpp}) and~(\ref{normalpp2}), we obtain
\begin{equation}
\int_0^\infty \mathrm{d}\omega~P_p (\omega)= m \omega_0^2 \alpha^{(0)}(0).
\end{equation} From Eq.~(\ref{sus1111}), it directly follows that putting \(\omega =0\) gives \(\alpha^{(0)}(0) = [m \omega_0^2]^{-1}\) which means \(P_p(\omega)\) is normalized. We refer the reader to \cite{kaur}, where the positivity and normalization of both \(P_k (\omega)\) and \(P_p(\omega)\) have been demonstrated in a more general setting where the number of spatial dimensions is three and there is an externally-applied magnetic field.

\section*{References}


\begin{thebibliography}{99}

\bibitem{brown}R. Brown, ``XXVII. A brief account of microscopical observations made in the months of June, July and August 1827, on the particles contained in the pollen of plants; and on the general existence of active molecules in organic and inorganic bodies", Phil. Mag. \textbf{4}, 161 (1828).

\bibitem{janI}J. Ingenhousz, ``Vermischte schriften physisch medicinischen inhalts", Wappler \textbf{2}, 123 (1784).

\bibitem{Hanggirev}J. Spiechowicz, I. G. Marchenko, P. H\"anggi, and J. \L{}uczka, ``Diffusion coefficient of a Brownian particle in equilibrium and nonequilibrium: Einstein
model and beyond'', Entropy \textbf{25}, 42 (2023).


\bibitem{brown1}A. Einstein, ``\"{U}ber die von der molekularkinetischen Theorie der W\"{a}rme geforderte Bewegung von in ruhenden Flussigkeiten suspendierten Teilchen", Ann. Phys. (Berl.)
\textbf{17}, 549 (1905).

\bibitem{brown2} M. Smoluchowski, ``Zur kinetischen Theorie der Brownschen Molekularbewegung und der Suspensionen", Ann. Phys. (Berl.) \textbf{21}, 756 (1906).

\bibitem{brown3} P. Langevin, ``Sur la theorie du mouvement brownien", Comptes rendus de l’Academie
des Sciences (Paris) \textbf{146}, 530 (1908).

\bibitem{brown4} J. Perrin, ``Les Atomes", F\'elix Alcan, Paris (1913).

\bibitem{brown5} G. E. Uhlenbeck and L. S. Ornstein, ``On the Theory of the Brownian Motion", Phys.
Rev. \textbf{36}, 823 (1930). Reprinted in ``Selected Papers on Noise and Stochastic Processes'', Dover Publications (2003).

\bibitem{kramer} H. A. Kramers, ``Brownian motion in a field of force and the diffusion model of chemical reactions", Physica \textbf{7}, 284 (1940).

\bibitem{lineshape} R. Kubo, ``A Stochastic Theory of Line Shape", Adv. Chem. Phys. \textbf{15}, 101 (1969).

\bibitem{johnson1} J. B. Johnson, ``Thermal Agitation of Electricity in Conductors", Nature \textbf{119}, 50 (1927).

\bibitem{johnson2} J. B. Johnson, ``Thermal Agitation of Electricity in Conductors", Phys. Rev. \textbf{32}, 97 (1928).

\bibitem{nyquist} H. Nyquist, ``Thermal Agitation of Electric Charge in Conductors", Phys. Rev. \textbf{32}, 110 (1928).

\bibitem{FV} R. P. Feynman and F. L. Vernon, ``The theory of a general quantum system interacting with a linear dissipative system", Ann. Phys. (N. Y.) \textbf{24}, 118 (1963).

\bibitem{CL} A. O. Caldeira and A. J. Leggett, ``Path integral approach to quantum Brownian motion", Physica A \textbf{121}, 587 (1983).

\bibitem{CL1} A. O. Caldeira and A. J. Leggett, ``Influence of Dissipation on Quantum Tunneling in Macroscopic Systems", Phys. Rev. Lett. \textbf{46}, 211 (1981).

\bibitem{CL2} A. O. Caldeira and A. J. Leggett, ``Quantum tunnelling in a dissipative system", Ann. Phys. (N.Y.) \textbf{149}, 374 (1983). 



\bibitem{lho1} G. W. Ford, M. Kac, and P. Mazur, ``Statistical Mechanics of Assemblies of Coupled Oscillators", J. Math. Phys. \textbf{6}, 504 (1965).

\bibitem{lho2} I. R. Senitzky, ``Dissipation in Quantum Mechanics. The Harmonic Oscillator", Phys. Rev. \textbf{119}, 670 (1960).

\bibitem{lho3} F. Schwabl and W. Thirring, ``Quantum theory of laser radiation", In: Ergebnisse der exakten Naturwissenschaften, vol 36. Springer (1964).

\bibitem{lho4} P. Ullersma, ``An exactly solvable model for Brownian motion", Physica \textbf{32}, 27 (1966).

\bibitem{lho5} F. Haake and R Reibold, ``Strong damping and low-temperature anomalies for the harmonic oscillator", Phys. Rev. A \textbf{32}, 2462 (1985).

\bibitem{lho6} G. W. Ford, J. T. Lewis, and R. F. O'Connel, ``Independent oscillator model of a heat bath: Exact diagonalization of the Hamiltonian", J. Stat. Phys. \textbf{53}, 439 (1988).
         
\bibitem{lho7} F. Haake and M. \.Zukowski, ``Classical motion of meter variables in the quantum theory of measurement", Phys. Rev. A \textbf{47},2506 (1993).

\bibitem{quantumreg}G. W. Ford and R. F. O’Connell, ``There is No Quantum Regression Theorem", Phys. Rev. Lett. \textbf{77}, 798 (1996).

\bibitem{fordphysicae2005} G. W. Ford and R. F. O'Connell, ``Entropy of a quantum oscillator coupled to a heat bath and implications for quantum thermodynamics", Physica E \textbf{29}, 82 (2005).

\bibitem{ocon1} R. F. O'Connell, ``Does the Third Law of Thermodynamics Hold in the Quantum Regime?", J. Stat. Phys. \textbf{124}, 15 (2006).

\bibitem{15} P. H\"anggi and G.-L. Ingold, ``Quantum Brownian motion and the Third Law of thermodynamics",
Acta Phys. Pol. B \textbf{37}, 1537 (2006). 

\bibitem{FordQT}G. W. Ford and R.F. O’Connell, ``Quantum thermodynamic functions for an oscillator coupled to a heat bath", Phys. Rev. B \textbf{75}, 134301 (2007).

\bibitem{hanggi}P. H\"anggi, G.-L. Ingold, and P. Talkner, ``Finite quantum dissipation: the challenge of obtaining specific heat", New J. Phys. \textbf{10}, 115008 (2008).

\bibitem{PRE79}J. Kumar, P. A. Sreeram, and S. Dattagupta, ``Low-temperature thermodynamics in the context of dissipative diamagnetism", Phys. Rev. E \textbf{79}, 021130 (2009).

\bibitem{specificheatingold} G.-L. Ingold, P. H\"anggi, and P. Talkner, ``Specific heat anomalies of open quantum systems", Phys. Rev. E \textbf{79}, 061105 (2009). 

\bibitem{sdg2}S. Dattagupta, J. Kumar, S. Sinha, and P. A. Sreeram, ``Dissipative quantum systems and the heat capacity", Phys. Rev. E \textbf{81}, 031136 (2010).

\bibitem{18}M. Bandyopadhyay, ``Quantum thermodynamics of a charged magneto-oscillator coupled to a heat bath", J. Stat. Mech. \textbf{2009}, P05002 (2009).

\bibitem{20} M. Bandyopadhyay, ``Dissipative Cyclotron Motion of a Charged Quantum-Oscillator and Third Law", J. Stat. Phys. \textbf{140}, 603 (2010).





\bibitem{jarzy1}P. Bialas and J. \L{}uczka, ``Kinetic Energy of a Free Quantum Brownian Particle",
Entropy \textbf{20}, 123 (2018).

\bibitem{jarzy2}J. Spiechowicz, P. Bialas, and J. \L{}uczka, ``Quantum partition of energy for a free Brownian particle: Impact of dissipation", Phys. Rev. A \textbf{98}, 052107 (2018).

\bibitem{jarzy3}P. Bialas, J. Spiechowicz, and J. \L{}uczka, ``Partition of energy for a dissipative quantum oscillator", Sci. Rep. \textbf{8}, 16080 (2018).

\bibitem{jarzy4}P. Bialas, J. Spiechowicz, and J. \L{}uczka, ``Quantum analogue of energy equipartition theorem", J. Phys. A: Math. Theor. \textbf{52}, 15LT01 (2019).

\bibitem{jarzy5} J. \L{}uczka, ``Quantum Counterpart of Classical Equipartition of Energy", J. Stat. Phys. \textbf{179}, 839 (2020).

\bibitem{jarzy6}J. Spiechowicz and J. \L{}uczka, ``Energy of a free Brownian particle coupled to thermal vacuum", Sci. Rep. \textbf{11}, 4088
(2021).

\bibitem{kaur}J. Kaur, A. Ghosh, and M. Bandyopadhyay, ``Quantum counterpart of energy equipartition theorem for a dissipative charged magneto-oscillator: Effect of dissipation, memory, and magnetic field", Phys. Rev. E \textbf{104}, 064112 (2021).



\bibitem{kaur2}J. Kaur, A. Ghosh, and M. Bandyopadhyay, ``Partition of free energy for a Brownian quantum oscillator: Effect of dissipation and magnetic field", Physica A  \textbf{599}, 127466 (2022).

\bibitem{kaur1}J. Kaur, A. Ghosh, and M. Bandyopadhyay, ``Quantum counterpart of energy equipartition theorem for fermionic systems", J. Stat. Mech. \textbf{2022}, 53105 (2022).

\bibitem{ghoshelec} A. Ghosh, ``Generalised energy equipartition in electrical circuits", Pramana \textbf{97}, 82 (2023).

\bibitem{kaur3}J. Kaur, A. Ghosh, and M. Bandyopadhyay, ``Partition of kinetic energy and magnetic moment in dissipative diamagnetism", Physica A \textbf{625}, 128993 (2023).

\bibitem{sdg1}S. Dattagupta and J. Singh, ``Landau Diamagnetism in a Dissipative and Confined System", Phys. Rev. Lett. \textbf{79}, 961 (1997).



\bibitem{Leg} A. J. Leggett, ``Quantum tunneling in the presence of an arbitrary linear dissipation mechanism", Phys. Rev. B \textbf{30}, 1208 (1984).

\bibitem{Andiss} A. Cuccoli, A. Fubini, V. Tognetti, and R. Vaia, ``Quantum thermodynamics of systems with anomalous dissipative coupling", Phys. Rev. E \textbf{64}, 066124 (2001).

\bibitem{Andiss0} Y. A. Makhnovskii and E. Pollak, ``Hamiltonian theory of stochastic acceleration", Phys. Rev. E \textbf{73}, 041105 (2006).

\bibitem{Andiss1} J. Ankerhold and E. Pollak, ``Dissipation can enhance quantum effects", Phys. Rev. E \textbf{75}, 041103 (2007).

\bibitem{momentum} S. Gupta and M. Bandyopadhyay, ``Quantum Langevin equation of a charged oscillator in a magnetic field and coupled to a heat bath through momentum variables", Phys. Rev. E \textbf{84}, 041133 (2010).

\bibitem{momentum1} S. Gupta and M. Bandyopadhyay, ``Free energy of a charged oscillator in a magnetic field and coupled to a heat bath through momentum variables", J. Stat. Mech. \textbf{2013}, P04034 (2013).



\bibitem{GaCoFT} G. Gallavotti and E. G. D. Cohen, ``Dynamical Ensembles in Nonequilibrium Statistical Mechanics", Phys. Rev. Lett. \textbf{74}, 2694 (1995).

\bibitem{JaFT} C. Jarzynski, ``Nonequilibrium Equality for Free Energy Differences", Phys. Rev. Lett. \textbf{78}, 2690 (1997).

\bibitem{JKFT} J. Kurchan, ``Fluctuation theorem for stochastic dynamics", J. Phys. A: Math. Gen. \textbf{31}, 3719 (1998).



\bibitem{iso1} N. Kumar, S. Ramaswamy, and A. K. Sood, ``Symmetry Properties of the Large-Deviation Function of the Velocity of a Self-Propelled Polar Particle", Phys. Rev. Lett. \textbf{106}, 118001 (2011).

\bibitem{iso2} P. I. Hurtedo, C. Perez-Espigares, J. J. del Pozo, and P. L. Garrido, ``Symmetries in fluctuations far from equilibrium", Proc. Natl. Acad. Sci. USA \textbf{108}, 7704 (2011).

\bibitem{iso3} N. Kumar, H. Soni, S. Ramaswamy, and A. K. Sood, ``Anisotropic isometric fluctuation relations in experiment and theory on a self-propelled rod", Phys. Rev. E. \textbf{91}, 030102(R) (2015).

\bibitem{GSA_SDG_PRE} G. S. Agarwal and S. Dattagupta, ``Generalized fluctuation theorems for classical systems", Phys. Rev. E \textbf{92}, 052139 (2015). 

\bibitem{GSA_SDG} G. S. Agarwal and S. Dattagupta, ``Quantum fluctuation theorem for dissipative cyclotron motion", arXiv:1601.05642.

 

\bibitem{meas1} S. Mukamel, ``Quantum Extension of the Jarzynski Relation: Analogy with Stochastic Dephasing", Phys. Rev. Lett. \textbf{90}, 170604 (2003).

\bibitem{meas2} M. Esposito, U. Harbola, and S. Mukamel, ``Nonequilibrium fluctuations, fluctuation theorems, and counting statistics in quantum systems", Rev. Mod. Phys. \textbf{81}, 1665 (2009).

\bibitem{meas3} P. Talkner, E. Lutz, and P. H\"anggi, ``Fluctuation theorems: Work is not an observable", Phys. Rev. E \textbf{75}, 050102(R) (2007).

\bibitem{meas4} M. Campisi, P. H\"anggi, and P. Talkner, ``Colloquium: Quantum fluctuation relations: Foundations and applications", Rev. Mod. Phys. \textbf{83}, 771 (2011).





\bibitem{ford1988}G. W. Ford, J. T. Lewis, and R. F. O’Connell, ``Quantum Langevin equation", Phys. Rev.
A \textbf{37}, 4419 (1988).



\bibitem{Weiss} U. Weiss, ``Quantum Dissipative Systems", World Scientific (1999).

\bibitem{ohmicdiverge} H. Grabert, P. Schramm, and G.-L. Ingold, ``Quantum Brownian motion: The functional integral approach", Phys. Rep. \textbf{168},
115 (1988).


\bibitem{HL}D. Boyanovsky and D. Jasnow, ``Heisenberg-Langevin
versus quantum master equation", Phys. Rev. A \textbf{96},
062108 (2017).


\bibitem{Tong:2023}X.-H. Tong, H. Gong, Y. Wang, R.-X. Xu, and Y. Yan, ``Multimode Brownian oscillators: Exact solutions to heat transport", J. Chem. Phys. \textbf{159}, 024117 (2023).

\bibitem{Das:2020}A. Das, A. Dhar, I. Santra, U. Satpathi, and S. Sinha, ``Quantum Brownian motion: Drude and Ohmic baths as continuum limits of the Rubin model", Phys. Rev. E \textbf{102}, 062130 (2020).

\bibitem{bez} W. Bez, ``Microscopic preparation and macroscopic motion of a Brownian particle", Z. Phys. B Condens. Matter \textbf{39}, 319 (1980).

\bibitem{physicaA} J.  S\'anchez-Ca\~nizares and F. Sols, ``Translational symmetry and microscopic preparation in oscillator models of quantum dissipation", Physica A \textbf{212}, 181 (1994).

\bibitem{hist} P. H\"anggi, ``Generalized langevin equations: A useful tool for the perplexed modeller of nonequilibrium fluctuations?", In: Concepts Of Stochastics And Kinetics,  Lecture Notes in Physics, Springer (2007).

\bibitem{purisdgbook} S. Dattagupta and S. Puri, ``Dissipative Phenomena in Condensed Matter: Some Applications", Springer (2004).




\bibitem{ford} G. W. Ford and R. F. O'Connell, ``Calculation of Correlation Functions in the Weak Coupling Approximation", Ann. Phys. (N. Y.) \textbf{276}, 144 (1999).


\bibitem{GSAQBM} G. S. Agarwal, ``Brownian Motion of a Quantum Oscillator", Phys. Rev. A \textbf{4}, 739 (1971).


\bibitem{Comm} P. de Smedt, D. D\"urr,  J. L. Lebowitz, and C. Liverani, ``Quantum system in contact with a thermal environment: Rigorous treatment of a simple model", Commun. Math. Phys. \textbf{120}, 195 (1988).

\bibitem{AG_SDG}A. Ghosh and S. Dattagupta, ``Weak-coupling limits of the quantum Langevin equation for an oscillator", arXiv:2404.01285.



\bibitem{virial} A. Ghosh and M. Bandyopadhyay, ``Quantum dissipation and the virial theorem", Physica A \textbf{625}, 128999 (2023).

\bibitem{agmb} A. Ghosh, J. Kaur, and M. Bandyopadhyay, ``Energetics of the dissipative quantum oscillator", Physica A \textbf{643}, 129782 (2024).


\bibitem{LBE}D. Chru\'sci\'nski and S. Pascazio, ``A Brief History of
the GKLS Equation", Open Sys. Inf. Dyn. 24, 1740001
(2017).



\bibitem{Boltzmann} L. Boltzmann, ``\"{U}ber die Natur der Gasmolek\"{u}le", Wiener Berichte \textbf{74}, 553 (1876).

\bibitem{waterson}J. J. Waterston and J. W. Strutt, ``I. On the physics of media that are composed of free and perfectly elastic molecules in a state of motion", Philos. Trans. R. Soc. A
\textbf{183}, 1 (1892).


\bibitem{reif} F. Reif, ``Fundamentals of Statistical and Thermal Physics", Waveland Press (2009).


\bibitem{landau} L. D. Landau and E. M. Lifshitz, ``Statistical Physics", 3rd ed., Butterworth-Heinemann (1996).

\bibitem{fordBB}G. W. Ford, J.T. Lewis, and R. F. O’Connell, ``Quantum oscillator in a blackbody radiation field II. Direct calculation of the energy using the fluctuation-dissipation theorem", Ann. Phys. (N. Y.) \textbf{185}, 270 (1988).

\bibitem{callenwelton} H. B. Callen and T. A. Welton, ``Irreversibility and Generalized Noise", Phys. Rev. \textbf{83}, 34 (1951).

\bibitem{Case}K. M. Case, ``On fluctuation-dissipation theorems", Trans. Theory Statist. Phys. \textbf{2}, 129 (1972).

\bibitem{Raphael1}R. Chetrite and S. Gupta, ``Two Refreshing Views of Fluctuation Theorems Through Kinematics Elements and Exponential Martingale", J. Stat. Phys. \textbf{143}, 543 (2011).


\bibitem{9}G. W. Ford, J. T. Lewis, and R. F. O’Connell, ``Quantum Oscillator in a Blackbody Radiation Field", Phys. Rev. Lett. \textbf{55}, 2273 (1985).




\bibitem{wilks} J. Wilks, ``The Third Law of Thermodynamics", Oxford University Press (1961).

\bibitem{myatt} C. J. Myatt, B. E. King, Q. A. Turchette, C. A. Sackett, D. Kielpinski, W. M. Itano, C. Monroe, and D. J. Wineland, ``Decoherence of quantum superpositions through coupling to engineered reservoirs", Nature \textbf{403}, 269 (2000).

\bibitem{capek} V. Capek and D. P. Sheehan, ``Challenges to The Second Law of Thermodynamics: Theory and Experiment", Springer (2005).


\bibitem{peirls} R. Peierls, ``Surprises in Theoretical Physics", Princeton University Press (1979).

\bibitem{magref1} L. Landau, ``Diamagnetismus der Metalle", Z. Phys. \textbf{64}, 629 (1930).

 \bibitem{singh}S. Dattagupta and J. Singh, ``Stochastic motion of a charged particle in a magnetic field: II Quantum Brownian treatment", Pramana \textbf{47}, 211 (1996).



\bibitem{22}J. H. van Vleck, ``The Theory of Electric and Magnetic Susceptibilities", Oxford University Press (1932).

\bibitem{darwin} C. G. Darwin, ``The Diamagnetism of the Free Electron", Math. Proc. Camb. Philos. Soc. \textbf{27}, 86 (1931).

\bibitem{cursci} S. Dattagupta, A. M. Jayannavar, and N. Kumar, ``Landau diamagnetism revisited", Curr. Sci. \textbf{80}, 861 (2001).
   
 
 
 
 \bibitem{balki} V. Balakrishnan, ``Elements of Nonequilibrium Statistical Mechanics", Springer Nature (2020).
 
 

\bibitem{SDBook} S. Dattagupta, ``Diffusion: Formalism and Applications", CRC Press (2013).
 
   
\bibitem{sdg-aharonov-bohm}S. Dattagupta, ``Quantum Phase and its Measurable Attributes à la Aharonov–Bohm Effect", Resonance \textbf{23}, 949 (2018).

\bibitem{deco1} H.D. Zeh, ``On the interpretation of measurement in quantum theory", Found. Phys. \textbf{1}, 69 (1970).

\bibitem{deco2}M. Schlosshauer, ``Decoherence, the measurement problem, and interpretations of quantum mechanics", Rev. Mod. Phys. \textbf{76}, 1267 (2005).

\bibitem{decorev}M. Schlosshauer, ``Quantum decoherence", Phys. Rep. \textbf{831}, 1 (2019).
 



\end{thebibliography}
\end{document}